\documentclass[aps,prb,nopacs,floatfix,12pt]{revtex4}

\usepackage{graphicx}
\usepackage{amssymb}
\usepackage{amsmath}
\usepackage{bm}
\usepackage{mathtools}
\usepackage{color}
\usepackage{makecell}   
\usepackage{mathrsfs}
\usepackage{amsfonts}
\usepackage{textcomp}
\usepackage{rotating}
\usepackage{comment}

\newcommand{\bra}{\langle}
\newcommand{\ket}{\rangle}

\newcommand{\Eph}{E_\mathrm{ph}}

\newcommand{\QQ}{\mathbf{Q}}
\newcommand{\RR}{\mathbf{R}}
\newcommand{\bmu}{\boldsymbol{\mu}}

\newcommand{\MatVar}[1]{\ensuremath{\underline{#1}}}%
\newcommand{\bfPsi}{\boldsymbol{\Psi}}

\begin{document}
\bibliographystyle{jcp}

\title{Photodissociation dynamics in the
first absorption band of pyrrole: II. Photofragment distributions for
the $^1\!A_2(\pi \sigma^*) \leftarrow \tilde{X}^1\!A_1(\pi\pi)$
transition}  
\author{David Picconi\footnote{Electronic mails:\,\,David.Picconi@ch.tum.de and david.picconi@gmail.com}}
\author{Sergy\ Yu.\ 
Grebenshchikov\footnote{Electronic mails: Sergy.Grebenshchikov@ch.tum.de and sgreben@gwdg.de}}
\affiliation{Department of Chemistry, Technical University of Munich,
  Lichtenbergstr. 4, 85747 Garching, Germany}

\begin{abstract}

The analysis of the total kinetic energy release (TKER) of the 
photofragments 
pyrrolyl + H-atom formed in the photodissociation of pyrrole in
the low-lying state $^1\!A_2(\pi \sigma^*)$ is presented. 
The TKER distributions contain 
complementary and often more precise information on the fragmentation
process than the broad diffuse absorption spectra. The distributions are 
calculated quantum mechanically  for the 
diabatic state $^1\!A_2(\pi \sigma^*)$ either isolated or coupled 
to the ground electronic state at an exit channel conical intersection. 
The calculations use 
the novel ab initio quasi-diabatic potential energy matrix constructed
in paper I. The approximate overlap integral-based 
adiabatic mapping approach is introduced with which the quantum mechanical
TKER distributions can be efficiently and accurately
reproduced. Finally, the  
calculated TKERs are compared with the experimental results.
The main features of the measured vibrationally resolved distributions 
are reproduced, and the spectral peaks are assigned and interpreted 
in detail.

\end{abstract}

\maketitle

\section{Introduction}
\label{intro}

This paper (paper II) continues the systematic analysis of the photochemistry
of pyrrole (C$_4$H$_4$NH) in the first absorption band started in 
Ref.\ \onlinecite{PG17A} (paper I). The absorption band is due to the 
two low lying $\pi\sigma^*$ states,
$^1\!A_2(\pi\sigma^*)$ and $^1\!B_1(\pi\sigma^*)$, which 
are repulsive along the
NH bond and form exit channel conical intersections (CIs) with the 
ground electronic state $\widetilde{X}^1\!A_1(\pi\pi)$ 
[Fig.\ \ref{adcurves}(a)]. The resulting topography
of the state intersections is typical of many model biochromophores, and this
study is motivated by the goal to clarify the impact of this topography on the
observable features of the photodynamics. Paper I introduced the
ab initio based full dimensional molecular Hamiltonian and analyzed the
absorption spectra corresponding to the ultraviolet (UV) excitation of the
first excited state $^1\!A_2(\pi\sigma^*)$ at wavelengths 
$\lambda \ge 240$\,nm. Paper II focuses on the rovibrational product state
and kinetic energy distributions 
of the pyrrolyl photofragment formed in its ground electronic state 
upon the light triggered H-atom elimination in the state 
$^1\!A_2(\pi\sigma^*)$. The preliminary results of the analysis of the 
CI $\tilde{X}/^1\!A_2(\pi\sigma^*)$ and its signatures in the product state
distributions, in particular in the photofragment excitation spectra, are
described in Ref.\ \onlinecite{GP17A}. 

The photofragment distributions for the wavelengths 
$\lambda$ between 254\,nm and 240\,nm, relevant for the present study, 
have been studied using the H-atom Rydberg tagging photofragment
translational spectroscopy.\cite{CNQA04,CDNA06,AKMNOS10} The final state
populations in these experiments are quantified in terms of the 
total kinetic energy release (TKER).  In 
the two-fragment channel ${\rm H} + {\rm pyrrolyl}$, the TKER spectrum 
for a given photolysis wavelength is uniquely related to the 
 rovibrational distribution of the pyrrolyl radical.\cite{CNQA04,CDNA06}
The observed kinetic energy distributions consist of two well resolved
components,\cite{CNQA04} a fast one
(average kinetic energy $E_{\rm kin} \ge 5000$\,cm$^{-1}$)
and a slow one ($E_{\rm kin} \sim 1000$\,cm$^{-1}$). 
Experimentalists
associate the fast component with the direct dissociation occurring on a
subpicosecond time scale of 20\,fs --- 
110\,fs,\cite{LRHR04,RWYCYUS13,WNSSAWS15}
 and the slow component
with the statistical decomposition involving the ground electronic state
and unfolding on the time scale of 1.1\,ps.\cite{LRHR04} 

In this paper, we give a detailed description of 
 the TKER spectra associated with the fast direct dissociation. In the
long wavelength tail of the first absorption band, the experimental TKER 
spectra consist of merely several peaks  indicating that only a limited number
of vibrational states of pyrrolyl is selectively populated in the 
reaction.\cite{CNQA04} Ashfold and co-workers considered this result as
an indication that pyrrole photodissociates in a vibrationally 
adiabatic fashion, 
with the modes orthogonal to the dissociation path preserving
their initial vibrational excitation during the reaction. This vibrationally
adiabatic mechanism was subsequently used to explain the TKER spectra 
observed in
other model biochromophores.\cite{ACDDN06} 

Several theoretical studies of the photofragment distributions in pyrrole
have been published. Domcke and co-workers performed a
quantum mechanical final state analysis using two- and
three dimensional diabatic
potentials and off-diagonal diabatic couplings.\cite{VLMSD05} The primary goal
was to investigate the dependence of the electronic branching ratios on the 
initial vibrational state of the parent molecule.
The formation of pyrrolyl was also studied using the
classical mechanical \lq direct dynamics' with all 24 degrees of freedom
included.\cite{MSMS15} The TKER and the angular distributions were 
calculated for the first four electronic states without the explicit  
construction of their potential energy surfaces, and the qualitative features
of the low resolution TKER distributions measured by 
Stavros and co-workers\cite{RWYCYUS13} were successfully reproduced. 

To date, the high resolution TKER distributions of Refs.\ 
\onlinecite{CNQA04} and \onlinecite{CDNA06} 
have been neither analyzed theoretically 
nor accurately assigned. Furthermore, the extent of the vibrational
adiabaticity in the photodissociation in the $\pi\sigma^*$ states
remains unknown. This gap is closed in the present study which complements 
the analysis of the absorption spectra in paper I with the quantum mechanical
calculations of the TKER distributions. 
The photofragment kinetic energy distributions measured by Ashfold and 
co-workers are of special significance for the theoretical studies: They are
the only experimental data sets with high frequency resolution, which are
available for the low lying $\pi\sigma^*$ states in pyrrole (the absorption 
cross sections could not be measured directly). These data sets serve as 
benchmarks for testing the quality of the constructed molecular Hamiltonian, 
the ab initio potential energy surfaces, and the quantum dynamics.
In parallel, approximate
quantum mechanical methods, which  
reduce the computational burden of the ab initio and the 
quantum dynamics steps, 
are introduced and tested. In paper I, the convolution approximation for the
absorption spectra has been implemented as a tool to calculate and to assign
the diffuse bands of the $\pi\sigma^*$ states. In paper II, we demonstrate that
the exact 
photofragment distributions can be reproduced using the adiabatic mapping
of the Franck-Condon (FC) excitations of the aromatic ring onto free fragment
states. Both approximations work reliably for pyrrole and are expected to be 
generally applicable to the fast direct dissociation  
in $\pi\sigma^*$ states.

The paper is organized as follows: The form of the molecular Hamiltonian
and the main results of paper I are 
recapitulated in Sect.\ \ref{pap1} in order to keep the presentation 
self-contained. The quantum mechanical approach to 
the photofragment distributions using the MCTDH method\cite{BJWM00}
and the overlap integral-based mapping approximation are
described in Sect.\ \ref{psd} and in the Appendix. In Sect.\ \ref{res_tker}, 
the TKER spectra are discussed for the calculations in which 
six, eleven, and fifteen internal degrees of freedom are included. These
are the same 6D, 11D, and 15D calculations, for which the absorption spectra
in paper I were analyzed and assigned. The emphasis here is on the dissociation
in the isolated state $1^1\!A_2(\pi\sigma^*)$. 
Conclusions are given  in Sect.\ \ref{sum}, complemented with a sketch of the
non-adiabatic effects observed in the two-state dissociation dynamics
of pyrrole in the electronic states  
$\tilde{X}/1^1\!A_2(\pi\sigma^*)$ weakly coupled at an exit channel
conical intersection. 


\section{Molecular Hamiltonian and Absorption Spectra}
\label{pap1}

\begin{figure}
\includegraphics[scale=0.1]{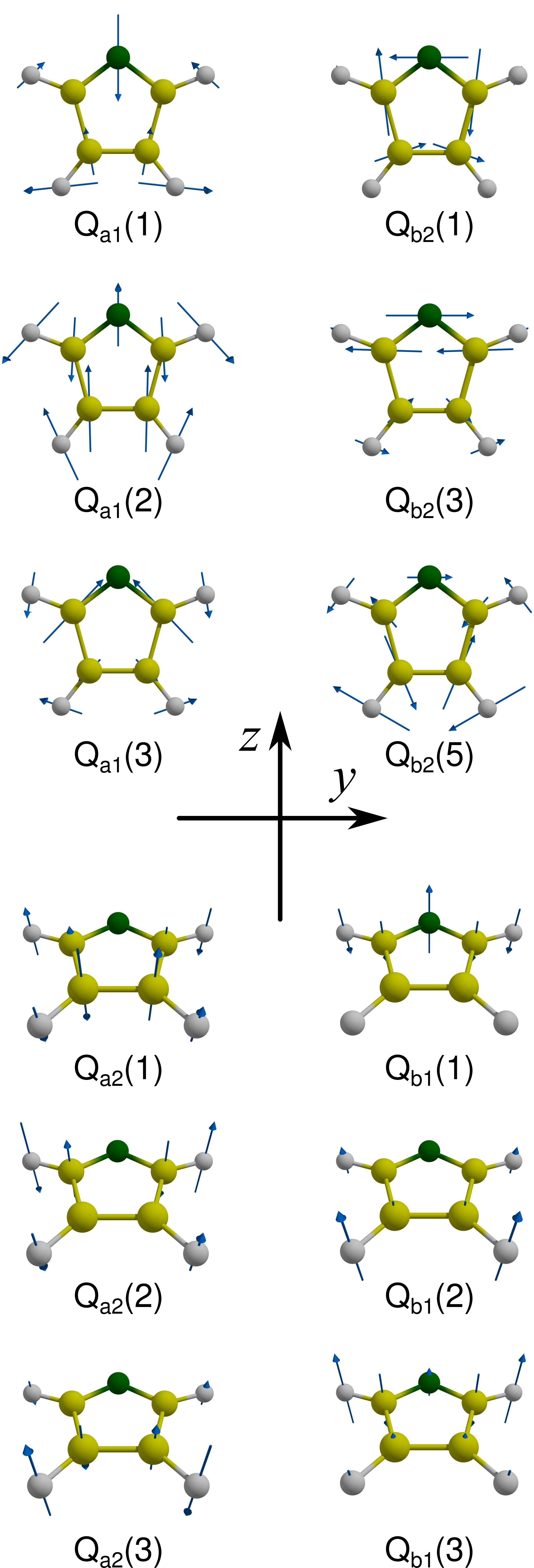} 
\caption{Sketches of the pyrrolyl normal modes belonging to all four irreps
of the C$_{\rm 2v}$ symmetry group. These modes are included in the 15D 
calculations described in Sect. \ref{res_tker_3}. Also shown is the pyrrole 
axis system in which the components $\mu^{A_2}_{x,y,z}$
of the TDM vector are set. The $x$ axis is perpendicular to the plane of the
ring.}
\label{pyrrolyl_modes}
\end{figure}

The molecular Hamiltonian, constructed in paper I for the 
electronic states $\widetilde{X}$, $A_2$, and $B_1$ in the locally diabatic
representation [see Fig.\ \ref{adcurves}(a)], is written as 
\begin{equation}
  \hat{\MatVar{{\bf H}}} = \hat{T}\,\MatVar{{\bf 1}} +
\left( \begin{array}{ccc} V^{X} & V^{X A_2} & V^{X B_1} \\
  V^{X A_2} & V^{A_2} & V^{A_2 B_1} \\
  V^{X B_1} & V^{A_2 B_1} & V^{B_1} \end{array} \right) \, ;
\label{Ham}
\end{equation}
bold faced underlined characters denote $3\times 3$ matrices. 
Pyrrole is described using (i) the \lq disappearing modes', i.e.
the three Jacobi coordinates $\RR \equiv (R,\theta,\phi)$ of the
dissociating H-atom defined relative to the center of mass of the pyrrolyl
fragment 
and (ii) the \lq non-disappearing modes', i.e. the 21
dimensionless normal modes $\QQ$ of the pyrrolyl ring. 
The normal modes $\QQ$ are partitioned into
four symmetry blocks $\QQ = \{\QQ_{a_1}, \QQ_{a_2}, \QQ_{b_1},
\QQ_{b_2} \}$. Examples of the normal modes in each symmetry block are given
in Fig.\ \ref{pyrrolyl_modes}.
 The kinetic energy operator in each diabatic state 
is written as a sum of the kinetic energies in the $\RR$- and $\QQ$-spaces, 
$\hat{T} = \hat{T}_R + \hat{T}_Q$, and the kinematic coupling between the
disappearing modes and the vibrations of the ring is neglected. 

The quasi-diabatic potential matrix $\MatVar{{\bf V}}$ 
in Eq.\ (\ref{Ham})
is constructed as a sum of two terms, 
\begin{equation}
\MatVar{{\bf V}}(\RR,\QQ) = \MatVar{{\bf U}}_R(\RR) +
\MatVar{{\bf W}}_Q(\QQ|R) \, .
\label{V_gen}
\end{equation}
The term  $\MatVar{{\bf U}}_R(\RR)$ depends
only on the disappearing modes $(R,\theta,\phi)$ and is constructed by spline
interpolations of the diabatized ab initio energies on a dense ab initio
three-dimensional (3D) grid in $\RR$.  The term 
$\MatVar{{\bf W}}_Q(\QQ|R)$ depends on 
the 21 non-disappearing modes $\QQ$ and on the interfragment
distance $R$. The form of the matrix $\MatVar{{\bf W}}_Q(\QQ|R)$
is similar to the vibronic coupling
model of Ref.\ \onlinecite{KDC84} (i.e. the pyrrolyl ring is treated using
quadratic Hamiltonians), but all model parameters are calculated with 
the CASPT2 method on a dense grid along $R$ and subsequently interpolated.
The off-diagonal diabatic matrix 
elements $V^{X A2}$ and $V^{X B_1}$ in Eq.\ (\ref{V_gen}) are kept localized 
along the tuning mode $R$ 
to the vicinities of the CIs between states they refer to. They vanish outside
the CI region (i.e. in the FC zone and in the dissociation channels) 
where the transition probability between states is negligible
and where the adiabatic and the diabatic states are constructed to coincide. 

The initial state of the parent molecule prior to photoexcitation is taken to 
be the ground vibrational state $\Psi_0(\RR,\QQ)$
in the potential $V^X$ of the ground 
electronic state $\widetilde{X}$. Immediately after the photoexcitation, the
molecule is in the state
\begin{equation}
\Phi_\epsilon(t=0) = \left( \bmu^{A_2}(\RR,\QQ)\cdot
\hat{\boldsymbol{\epsilon}}\right) \Psi_0(\RR,\QQ) \, ,
\label{InitState2}
\end{equation}
where $\hat{\boldsymbol{\epsilon}} = x, y$ or $z$ denotes the polarization
vector of the electric field of the incident light and $\bmu^{A_2}(\RR,\QQ)$ 
is the transition dipole moment (TDM) vector of the state 
$1^1\!A_2(\pi\sigma^*)$ with $\widetilde{X}$. The molecular axes are chosen so
that $z$ runs along the NH bond, $y$ lies in the ring plane, $x$ is 
perpendicular to the ring plane, and all three are mutually orthogonal 
(see Fig. \ref{pyrrolyl_modes}). 

The state $1^1\!A_2(\pi\sigma^*)$ is optically
dark, and the transition $1^1\!A_2(\pi\sigma^*) \leftarrow \widetilde{X}$
becomes vibronically allowed upon displacements along the symmetry breaking
modes of $b_1$, $b_2$, and $a_2$ symmetry. For this reason, the coordinate
dependence of the TDMs in Eq.\ (\ref{InitState2}) is explicitly taken into
account: This dependence materially affects the observed absorption
spectra and the product state distributions in the $\pi\sigma^*$ states. 
The coordinate dependent TDM vector functions $\bmu^{A_2}(\RR,\QQ)$ are
written in the form
\begin{equation}
\bmu^{A_2}(\RR,\QQ) \approx \bmu_R^{A_2}(\RR) + \bmu_Q^{A_2}(\QQ|R) \, .
\label{tdm0}
\end{equation}
consistent with the well-known Herzberg-Teller expansion.\cite{BUNKERJENSEN06}
Two realizations of Eq.\ (\ref{tdm0}) are considered. One of them has
the standard Herzberg-Teller form and includes only small deviations 
from the FC geometry, so that the  TDM functions are 
strictly separable in $\RR$ and $\QQ$: 
\begin{subequations}
\label{muHT}
\begin{align}
\mu^{A_2}_x(\RR,\QQ) & = & \mu^{A_2}_{x,\theta}(R_{\rm FC})
\sin(\theta) \sin(\phi) +
\sum_{i} \mu^{A_2}_{x,i}(R_{\rm FC})  Q_{b_2}(i) \  ,
\label{mu_x_A2} \\
\mu^{A_2}_y(\RR,\QQ) & = &  \mu^{A_2}_{y,\theta}(R_{\rm FC})
\sin(\theta) \cos(\phi) + \sum_{i} \mu^{A_2}_{y,i}(R_{\rm FC})
Q_{b_1}(i) \ ,
\label{mu_y_A2} \\
\mu^{A_2}_z(\RR,\QQ) & = &  \sum_{i} \mu^{A_2}_{z,i}(R_{\rm FC})
Q_{a_2}(i) \, .
\label{mu_z_A2}
\end{align}
\end{subequations}
The sums run over all ring normal modes $Q_\Gamma(i)$ of the indicated 
symmetry $\Gamma$. The dependence on the disappearing angles is 
expressed in terms of the real spherical harmonics
$p_x$ and $p_y$. 
The intensity of excitation of a particular ring vibration $Q(i)$ in 
the state $1^1\!A_2(\pi\sigma^*)$ is proportional to the square of the 
Herzberg-Teller coefficient $\mu^{A_2}_{\epsilon,i}$. The numerical values
of the coefficients are given in Table III of paper I. 
These TDMs  were used in 
paper I in the calculations of the 6D, 11D, and 15D absorption spectra 
of the state $1^1\!A_2(\pi\sigma^*)$ [see Fig.\ \ref{spec}(a---c)]. The
analysis of the respective TKER distributions is given in 
Sect.\ \ref{res_tker}. 

A slightly different set  
of functions, which allows for a more realistic representation of the 
coordinate dependence of ab initio TDMs, 
is used to calculate TKER distributions which are compared 
with the experimental data of Ref.\ \onlinecite{CNQA04}: 
\begin{subequations}
\label{dipA2_R}
\begin{align}
\mu^{A_2}_x(\RR,\QQ) & = & \mu^{A_2}_{x,\theta,1}(R) \sin \theta \sin \phi +  
\mu^{A_2}_{x,\theta,2}(R) \sin(2\theta) \sin \phi +
\sum_{i} \mu^{A_2}_{x,i}(R)  Q_{b_2}(i) \  ,  
\label{mu_x_A2_R} \\
\mu^{A_2}_y(\RR,\QQ) & = & \mu^{A_2}_{y,\theta,1}(R) \sin \theta \cos \phi + 
\mu^{A_2}_{y,\theta,2}(R) \sin(2\theta) \cos \phi
 + \sum_{i} \mu^{A_2}_{y,i}(R)  Q_{b_1}(i) \ ,  
\label{mu_y_A2_R} \\
\mu^{A_2}_z(\RR,\QQ) & = & \sum_{i} 
\mu^{A_2}_{z,i}(R) Q_{a_2}(i) \, .
\label{mu_z_A2_R}
\end{align}
\end{subequations}
These functions are similar to the Herzberg-Teller expression of 
Eq.\ (\ref{muHT}), with the following differences. First, 
the real spherical harmonics $d_{xz}$ and $d_{yz}$ are added to  
$p_x$ and $p_y$ to better describe the angular dependence of the TDMs. 
Second, the Herzberg-Teller coefficients are made 
$R$ dependent via the second order Taylor expansion around the FC point 
$R_\mathrm{FC}$:
\begin{equation}
\mu(R) = \mu^{(0)}(R_{\rm FC}) + \mu^{(1)}(R_{\rm FC}) 
\left(R - R_{\rm FC}\right) + \
\mu^{(2)}(R_{\rm FC}) \left(R - R_{\rm FC}\right)^2 \ . 
\label{mu_R}
\end{equation}
Finally, only the modes making the largest contributions to the TDM are kept
in the Eq.\ (\ref{dipA2_R}). The expansion coefficients, found by fitting 
to the ab initio data, are summarized in Table \ref{T00}. 

The photon energy dependent total absorption cross section is an incoherent
average over the different electric field orientations, 
\begin{equation}
\sigma_{\rm tot}(\omega) =
\frac{1}{3} \sum_{\epsilon=x,y,z} \sigma_\epsilon(E_{\rm ph}) \, .
\label{spectrum2}
\end{equation}
Each cross section $\sigma_\epsilon(E_{\rm ph})$ can be expressed as an 
expectation value\cite{PG15}
\begin{equation}
\label{totCS_En}
\sigma_\epsilon(E_\mathrm{ph}) =  \frac{E_\mathrm{ph}}{3c \epsilon_0} 
\lim_{\lambda \rightarrow 0} \lambda \left\bra 
\bfPsi_\epsilon^\lambda (\QQ|\Eph) | 
\bfPsi_\epsilon^\lambda(\QQ|\Eph) \right\ket \ ,
\end{equation}
where the wave function
$\Psi_\epsilon^\lambda(E_{\rm ph})$ is a stationary energy component of
$\Phi_\epsilon(t=0)$:
\begin{equation}
\label{cross2}
\Psi_\epsilon^\lambda(E_{\rm ph}) = \hat{G}^+(E_{\rm ph})
\Phi_\epsilon(t=0)  \, . 
\end{equation}
Here $\hat{G}^+(E)$ is the advanced Green's function,
\begin{equation}
\label{green}
\hat{G}^+(E_{\rm ph}) =
-i\int_0^\infty e^{-i \left(\hat{H} - i \lambda \right)t}
e^{i E_{\rm ph} t} dt \, ,
\end{equation}
for the state $1^1\!A_2(\pi\sigma^*)$  
and $\lambda$ in the above equations signifies the absorbing potential 
introduced at the edge of the coordinate grid. The quantum mechanical 
absorption spectra used in this work are shown
in Fig.\ \ref{spec}. They have smooth absorption envelopes modulated by
the diffuse bands associated with excitations of the vibrational 
states in the shallow minimum of the state $1^1\!A_2(\pi\sigma^*)$ 
in the FC zone.

The quantum mechanical absorption spectrum for any polarization
$\epsilon$ can be approximated
by a convolution of the contributions due to the departing H-atom 
(in the space $\RR$ of the disappearing modes)
and due to the pyrrolyl ring (in the space $\QQ$ of the ring modes).
The convolution approximation was developed in paper I. 
First, the initial state of the parent molecule
$\Psi_0(\RR,\QQ)$ was
replaced by a product of an $\RR$- and a $\QQ$-dependent factor,
$\Psi_0(\RR,\QQ) \approx \Psi_R(\RR) \Psi_Q(\QQ)$. With the 
Herzberg-Teller TDMs, the 
photoexcited state $\Phi_\epsilon(t=0)$ was in the product form
\begin{equation}
\Phi_\epsilon(0) \approx F_R(\RR)f_Q(\QQ) \, ,
\label{InitState3}
\end{equation}
or [e.g. in the case of the $R$-dependent Herzberg-Teller coefficients of
Eq.\ (\ref{dipA2_R})] in the sum-of-products form:
\begin{equation}
\Phi_\epsilon(0) \approx \sum F_R(\RR)f_Q(\QQ) \, .
\label{InitState3a}
\end{equation}
Next, the diabatic potential of the state $1^1\!A_2(\pi\sigma^*)$, 
Eq.\ (\ref{V_gen}),  was simplified into a sum of 
purely $\RR$- and $\QQ$-dependent terms,
$V^{A_2} \approx U_R^{A_2}(\RR) + W^{A_2}_Q(\QQ|R_{\rm FC})$, 
with the $\QQ$-space term evaluated at a dissociation coordinate $R$
fixed near $R_{\rm FC}$ 
in the FC zone. The molecular Hamiltonian was approximated by a 
separable expression: 
\begin{equation}
\hat{H}_0 = \hat{T}_R + U_R^{A_2}(\RR) + 
\hat{T}_Q + W^{A_2}_Q(\QQ|R_{\rm FC}) \equiv \hat{H}_R(\RR) + 
\hat{H}_Q(\QQ|R_{\rm FC}) \, .
\label{V_gen2}
\end{equation}

With these two approximations, the absorption cross section assumed the
form 
\begin{equation}
\sigma(E_{\rm ph}) =
\frac{E_{\rm ph}}{2\epsilon_0 c} \int_{-\infty}^\infty
\bar{\sigma}_R(E_{\rm ph}-\omega)\bar{\sigma}_Q(\omega)
{\rm d}\omega \, ,
\label{spectrum3}
\end{equation}
where the \lq spectral functions' 
$\bar{\sigma}_R$ and $\bar{\sigma}_Q$ (i.e. 
the cross sections without energy
prefactors) were calculated separately
for the disappearing modes and for the pyrrolyl ring. The function 
$\bar{\sigma}_R$
is a smooth almost structureless envelope corresponding to the fast 
direct dissociation in the repulsive $\pi\sigma^*$ state. The function 
$\bar{\sigma}_Q$ consists of
a series of $\delta$-peaks of vibrational excitations in a
bound-bound transition. For
the vibronic coupling model, in which the Hamiltonian of the heteroaromatic
ring is represented as a set of coupled harmonic oscillators,  the
factor 
$\bar{\sigma}_Q(E)$ is simply a FC spectrum. If the photoexcited state has
a simple product form of Eq.\ (\ref{InitState3}), the spectrum 
$\bar{\sigma}_Q(E)$ is given by the overlap integrals,
\begin{equation}
\bar{\sigma}_Q(E) = \sum_{\bm m}
\left|\langle \varphi_{\bm m}(\QQ)|f_Q(\QQ)\rangle\right|^2
\delta(E-E_{\bm m})
\, ,
\label{spectrum4}
\end{equation}
between the eigenfunctions $\varphi_{\bm m}(\QQ)$ 
of the ring modes in the FC zone (with energies
$E_{\bm m}$ and the vector of the quantum numbers 
${\bm m}$),
and the initial state $f_Q$. For the sum-of-products form of Eq.\
(\ref{InitState3a}), the final expressions, derived in paper I, are 
slightly more involved.  

The convolution approximation  for the H-atom elimination reactions
substantially reduces the computational effort required 
to construct the molecular Hamiltonian and to calculate the
absorption spectrum. Indeed, the explicit 
construction of the potential
energy surfaces and the numerical solution of the nuclear Schr\"odinger
equation are restricted to the space of the three disappearing modes 
regardless of the actual size of the biochromophore. 
The good 
accuracy of the convolution approximation was demonstrated in paper I. 
It is also a convenient starting point for the overlap integral-based
approach to the photofragment distributions discussed in Sect.\ \ref{map1}.

\section{Photofragment distributions}
\label{psd}

\subsection{Quantum mechanical calculations}
\label{qm2}

The quantum mechanical calculations of the rovibrational photofragment 
distributions in the electronic channel ${\rm H + pyrrolyl}(^2\!A_2)$, 
diabatically correlating with the state $1^1\!A_2(\pi\sigma^*)$, 
are performed using the projection method of Balint-Kurti and 
coworkers,\cite{BDM90,BK04} which is formulated here 
in the time-independent framework.\cite{MA92,ARM92,PG15} 
The partial photodissociation cross section for the formation of pyrrolyl 
in a final vibrational state ${\bm n}$ is 
given by:\cite{BK04,PG15}
\begin{equation}
\label{cross1}
\sigma(E_{\rm ph},{\bm n}) = 
\frac{E_{\rm ph}}{3c\epsilon_0} 
\lim_{\lambda \rightarrow 0}\lambda\left|
\langle \psi^-_{\bm n}|\Psi^\lambda(E_{\rm ph})\rangle
\right|^2 \, = \frac{\Eph}{3c\epsilon_0} 
\left|T_{\bm n}(E_{\rm ph})\right|^2 \, ;
\end{equation}
the index $\epsilon$ indicating the polarization direction of the
incident light is omitted; 
the rotations of pyrrolyl and the quantum numbers of the associated coordinates
$\theta$ and $\phi$ are not explicitly included here, although
they are taken into account in the actual calculations.\cite{NOTE-PYR01B-01}
The scattering state $\psi_{\bm n}^-(E_{\rm ph})$
in the dissociation continuum describes the atom and the radical with photon
energy $E_{\rm ph}$ recoiling 
into the channel ${\bm n}$.  The energy resolved state 
$\Psi^\lambda(E_{\rm ph})$ is defined in Eq.\ (\ref{cross2}). 
At large pyrrolyl---H distances $R \rightarrow \infty$, it
contains purely outgoing waves along $R$:
\begin{equation}
\label{cross3}
\Psi^\lambda(E_{\rm ph}) \xrightarrow{R \rightarrow \infty} -\sum_{\bm n} 
T_{\bm n}(E_{\rm ph})\sqrt{\frac{m_R}{k_{\bm n}}}\,e^{ik_{\bm n}R}
\chi_{\bm n}(\QQ)
\, .
\end{equation}
The wave functions
  $\chi_{\bm n}(\QQ)$ are the vibrational eigenstates
of pyrrolyl with energies $E_{\bm n}$ and 
$k_{\bf n} = \sqrt{2m_R(E_{\rm ph}-E_{\bm n})}$ is the 
channel momentum;\cite{NOTE-PYR01B-01} $m_R \approx m_H$ is the pyrrolyl/H
reduced mass.  
The amplitudes in each channel are the photodissociation matrix elements 
 $T_{\bm n}(E_{\rm ph})$ which thus contain the 
dynamical information on the dissociation process.\cite{BK04} The $T$-matrix
elements are found by introducing the projection operators 
\begin{equation}
\label{proj1}
\hat{\cal{P}}_{\bm n} = 
\delta\left(R-R_\infty\right)|\chi_{\bm n}(\QQ)\rangle
\, .
\end{equation}
The operators $\hat{\cal{P}}_{\bm n}$ act in the $\QQ$-space
and project onto $\chi_{\bm n}$ at the analysis line $R = R_\infty$ located 
in the asymptotic region. An application to the state 
$\Psi^\lambda(E)$ gives: 
\begin{equation}
\label{cross4}
\left|T_{\bm n}(E_{\rm ph})\right|^2 \sim \frac{k_{\bm n}}{m_R}
\left|\langle\hat{\cal{P}}_{\bm n}^*|\Psi^\lambda(E_{\rm ph})\rangle\right|^2
\, .
\end{equation}
The matrix elements $\left|T_{\bm n}(E_{\rm ph})\right|^2$ are, with a 
proper normalization, the vibrational photofragment distributions. 
Summation of the partial cross sections 
over all quantum numbers ${\bm n}$ gives the total absorption cross section
of Eq.\ (\ref{totCS_En}). The TKER spectrum\cite{AKMNOS10,CNQA04} 
$I_{\rm TKER}(E_{\rm kin}|E_{\rm ph})$ is obtained via transforming the
internal energy distributions to the photofragment
kinetic energy scale:
\begin{equation} 
I_{\rm TKER}(E_{\rm kin}|E_{\rm ph}) = \sum_{\bm n}
\left|T_{\bm n}(\Eph) \right|^2 
\delta\left(\Eph - E_{\bm n} - E_\mathrm{kin} \right) \ .
\label{tker1}
\end{equation}
Thus, each peak in the TKER spectrum corresponds to a vibrational state of
the pyrrolyl fragment populated during photodissociation. 
The vibrational energies $E_{\bm n}$ in the above expression
are defined relative to the 
ground vibrational state of the ground electronic state $\widetilde{X}$; 
in this scale, the energy of the
lowest channel $E_0$ is exactly the 
quantum mechanical dissociation threshold $D_0$.  
The practical calculation of the photofragment distributions using the MCTDH
program package is outlined in the Appendix.

\subsection{Overlap integral-based mapping calculations of the
photofragment distributions}
\label{map1}

The $T$-matrix elements and the photofragment distributions
can be calculated approximately by means of
the overlap integral-based mapping. The approach can be illustrated using
the convolution method for the absorption spectra, 
introduced in paper I and sketched in Sect.\ \ref{pap1}. 
The convolution of the spectral
functions in Eq.\ (\ref{spectrum3}), is valid for the Green's functions
as well:\cite{PERELOMOV98}
\begin{equation}
\label{green2}
\hat{G}^+(E_{\rm ph}) \approx \hat{G}_0^+(E_{\rm ph}) = 
-\left(2\pi i\right)^{-1}
\int_{-\infty}^\infty \hat{G}_R^+(E_{\rm ph}-\omega) 
\hat{G}_Q^+(\omega)\,d\omega 
\, .
\end{equation}
The Green's function 
 $\hat{G}_R^+(E) = (E-\hat{H}_R(\RR)+i\lambda)^{-1}$ refers to the 
$\RR$ space of three
disappearing modes.   The vibrational spectrum of the 
pyrrolyl normal modes is discrete [cf. Eq.\ (\ref{spectrum4})], and 
the $\QQ$ space Green's function 
$\hat{G}_Q^+(E) = (E-\hat{H}_Q(\QQ|R_{\rm FC})+i\lambda)^{-1}$ can be 
represented as a sum over the vibrational states. The convolution
integral can therefore be rewritten as
\begin{equation}
\label{green3}
\hat{G}_0^+(E_{\rm ph}) = \sum_{\bm m} \hat{G}_R^+(\Eph-E_{\bm m}) 
|\varphi_{\bm m}\rangle\langle\varphi_{\bm m}|
\, ,
\end{equation}
where index ${\bm m}$ numbers the eigenstates $\varphi_{\bm m}$ of the
ring in the FC zone. 
Applying the Green's function $\hat{G}_0^+$ to the initial state
$\Phi(t=0)$ gives an approximation to the stationary energy component 
$\Psi^\lambda(E_{\rm ph})$. For the initial state in the simple product
form, $\Phi(0) = F_Rf_Q$ [cf. Eq.\ (\ref{InitState3})], one finds
\begin{equation}
\label{cross5}
\Psi^\lambda(E_{\rm ph}) \approx \hat{G}_0^+(E_{\rm ph})\Phi(0)
= \sum_{\bm m} 
\left[\hat{G}_R^+(\Eph-E_{\bm m})F_R(\RR)\right]
\langle \varphi_{\bm m}|f_Q\rangle \varphi_{\bm m}(\QQ)
\, .
\end{equation}
The Green's function $\left[\hat{G}^+_RF_R\right]$
acting on the initial state in the $\RR$ space generates outgoing
waves along $R$ for each vibrational state of the ring 
initially excited with the amplitude 
$\langle \varphi_{\bm m}|f_Q\rangle$. 
Although the above derivation uses 
separability of the Hamiltonian near $R_{\rm FC}$, 
the form of $\Psi^\lambda(E_{\rm ph})$, in particular the expansion in 
the ring states, is approximately valid for larger  $R$ along the 
dissociation path. The requirement is that 
the eigenstates $\varphi_{\bm m}(\QQ)$ vary
smoothly along the dissociation coordinate and commute with the 
kinetic energy $\hat{T}_R$. This implies an adiabatic evolution in the 
coordinates orthogonal to the reaction path:  
As interfragment distance grows,
the states $\varphi_{\bm m}$ with energies $E_{\bm m}$ very 
gradually turn into 
the vibrational states of the free pyrrolyl $\chi_{\bm n}$ with energies
$E_{\bm n}$, 
$(E_{\bm m},\varphi_{\bm m})
\xrightarrow{R\rightarrow\infty} (E_{\bm n},\chi_{\bm n})$.
For large $R$, the adiabatic  
state $\Psi^\lambda(E_{\rm ph})$ of Eq.\ (\ref{cross5}) 
is in the channel form of Eq.\ (\ref{cross3}): 
\begin{equation}
\label{cross6}
\Psi^\lambda(E_{\rm ph}) \xrightarrow{R\rightarrow\infty} -\sum_{\bm n} 
\left[\bar{\sigma}_R(\Eph-E_{\bm n})\right]^{1/2}e^{i\alpha_R}
\langle \varphi_{\bm m}|f_Q\rangle
\sqrt{\frac{m_R}{k_{\bm n}}}\,e^{ik_{\bm n}R}
\chi_{\bm n}(\QQ)
\, ,
\end{equation}
where $[\bar{\sigma}_R(\Eph-E_{\bm n})]^{1/2}\,e^{i\alpha_R}$
stands for the complex amplitude of the outgoing wave of 
$\left[\hat{G}^+_RF_R\right]$ along the reaction coordinate in a given
channel ${\bm n}$. 
Application of the projector $\hat{\cal{P}}_{\bm n}$ gives 
the vibrational state distributions in the mapping approximation: 
\begin{equation}
\label{cross7}
\left|t_{\bm n}(E_{\rm ph})\right|^2 \sim \frac{k_{\bm n}}{m_R}
\left|\langle \varphi_{\bm m}(\QQ)|f_Q(\QQ)\rangle\right|^2
\bar{\sigma}_R(\Eph-E_{\bm n})
\, .
\end{equation}
They are proportional to the FC overlap integrals in the $\QQ$ space, taken at
the excitation point $R_{\rm FC}$ and weighted with the \lq radial factor'
$\bar{\sigma}_R = \lim_{\lambda \rightarrow 0}\lambda 
\langle \hat{G}_R^+F_R|\hat{G}_R^+F_R\rangle$. 
The physical interpretation in view of
Eq.\ (\ref{spectrum4}) is that the population of a given product state
$\chi_{\bm n}$ is controlled by the intensity 
$\left|\langle \varphi_{\bm m}|f_Q\rangle\right|^2$ of excitation of the 
adiabatically connected state $\varphi_{\bm m}$ in the FC zone, multiplied
by the probability $\bar{\sigma}_R(\Eph-E_{\bm n})$ 
of excitation of the radial dissociative motion with the translational energy
$E_{\rm kin} = \Eph - E_{\bm n}$.\cite{NOTE-PYR01B-00} In other words, 
the harmonic populations of the
non-disappearing modes in the FC zone are adiabatically translated
to the infinite interfragment separation and 
mapped onto adiabatically connected product states. 
In the adiabatic
regime, the dissociation dynamics in the $\RR$ space unfolds on the 
adiabatic potentials $E_{\bm m}(\RR)$. These are shown in 
Fig.\ \ref{adcurves}(f) for the ring modes of $a_1$ symmetry. 
They are constructed by diagonalizing the Hessian
matrix of the ring modes for a set of $R$ values, and calculating the
harmonic energies of one-quantum excitations in each mode. 
The adiabatic curves are almost strictly 
parallel to each other, implying small non-adiabatic derivative couplings
and the prevailing 
adiabatic (i.e. the  \lq Born-Oppenheimer') evolution along the 
vibrational curves. Several avoided crossings between the curves
mark NH distances at which the residual vibrational energy exchange takes
place. 

In the actual application of the adiabatic mapping approximation, 
each normal mode in the FC region is associated with the pyrrolyl mode 
with the largest squared Duschinsky overlap. In this way, 
a one-to-one mapping is defined between the 
quantum numbers ${\bm m}$ of the FC modes and the quantum
numbers ${\bm n}$ of the free pyrrolyl modes (which is equivalent to the
mapping between the wavefunctions
$\varphi_{\bm m} \rightarrow \chi_{\bm n}$). 

The expression for the product state distributions in the 
overlap integral-based adiabatic mapping looks similar to the semiclassical 
FC mapping expression familiar in the context of the triatomic 
photodissociation (see, for example, Refs. 
\onlinecite{CS83A,BK86,HASCH87}). The FC mapping is recovered if the 
spectral amplitudes $\langle\varphi_{\bm m}|f_Q\rangle$ are replaced with the
projections $\langle\chi_{\bm n}|f_Q\rangle$ of the initial wave function 
directly onto the asymptotic product states. Test calculations demonstrated,
however, that the FC mapping is not accurate for pyrrole. 

The adiabatically mapped 
final state distributions can also be 
constructed for an initial
state of Eq.\ (\ref{InitState3a}), i.e. for  
$\Phi(0) = \sum F_Rf_Q$. Now several terms $\left[\hat{G}^+_RF_R\right]$
generate outgoing waves
with amplitudes $[\bar{\sigma}_R(\Eph-E_{\bm n})]^{1/2}\,e^{i\alpha_R}$, 
and the $T$-matrix elements are given by
\begin{equation}
\label{cross8}
\left|t_{\bm n}(E_{\rm ph})\right|^2 \sim \frac{k_{\bm n}}{m_R}
\left|\sum \langle \varphi_{\bm m}(\QQ)|f_Q(\QQ)\rangle
\bar{\sigma}_R(\Eph-E_{\bm n})^{1/2}\,e^{i\alpha_R}\right|^2
\, .
\end{equation}
Although the form of Eq.\ (\ref{cross8}) is less transparent, the 
calculation remains straightforward.
The complex amplitudes for each $F_R$ can be reconstructed, for example,
using the projection method of Balint-Kurti applied in the dissociative
$\RR$ space. 

Equations (\ref{cross7}) and (\ref{cross8}) are the main result of this
section. The actual implementation of them involves three main steps: First, 
the spectral function  $\bar{\sigma}_R$ (or the wave 
amplitudes $[\cdots]^{1/2}\,e^{i\alpha_R}$) 
are calculated by solving the 3D Schr\"odinger equation in the $\RR$ space. 
Second, the overlap integrals
$\langle \varphi_{\bm m}(\QQ)|f_Q(\QQ)\rangle$ are determined for the
ring modes in the FC zone. Finally, 
the mapping between the modes in the FC zone and in the
asymptotic region, $\varphi_{\bm m} \rightarrow \chi_{\bm n}$,  
is established. With the input produced in these three steps,
the $T$-matrix elements 
$\left|t_{\bm n}(E_{\rm ph})\right|^2$ are found, 
and the TKER spectrum is calculated using Eq.\ (\ref{tker1}).


\section{Results}
\label{res_tker}

\subsection{Quantum mechanics versus adiabatic mapping}
\label{qm-adia}

The TKER spectra of the recoiling C$_4$H$_4$N and H  
are evaluated for the 
6D, 11D, and 15D absorption spectra. The spectra calculated using the
Herzberg-Teller TDMs of Eq.\ (\ref{muHT}) are shown
in Fig.\ \ref{spec}. The 6D and 11D calculations are performed for the isolated
electronic state $1^1\!A_2$, with the 
non-totally symmetric modes  $b_1$ (for 6D) and all 
totally symmetric modes $a_1$ (for 11D) dynamically active.   
The 15D calculation is performed for the coupled pair $\widetilde{X}/A_2$. It
includes all modes of $a_2$ and $b_1$ symmetry, three $a_1$ modes with the
largest displacement between the minima of pyrrole and pyrrolyl, as well
as three $b_2$ modes along which the Herzberg-Teller coefficients of the
TDMs are the largest. The parameters of all MCTDH
calculations are summarized in Table VI of paper I.

\begin{figure}
\includegraphics[scale=0.33]{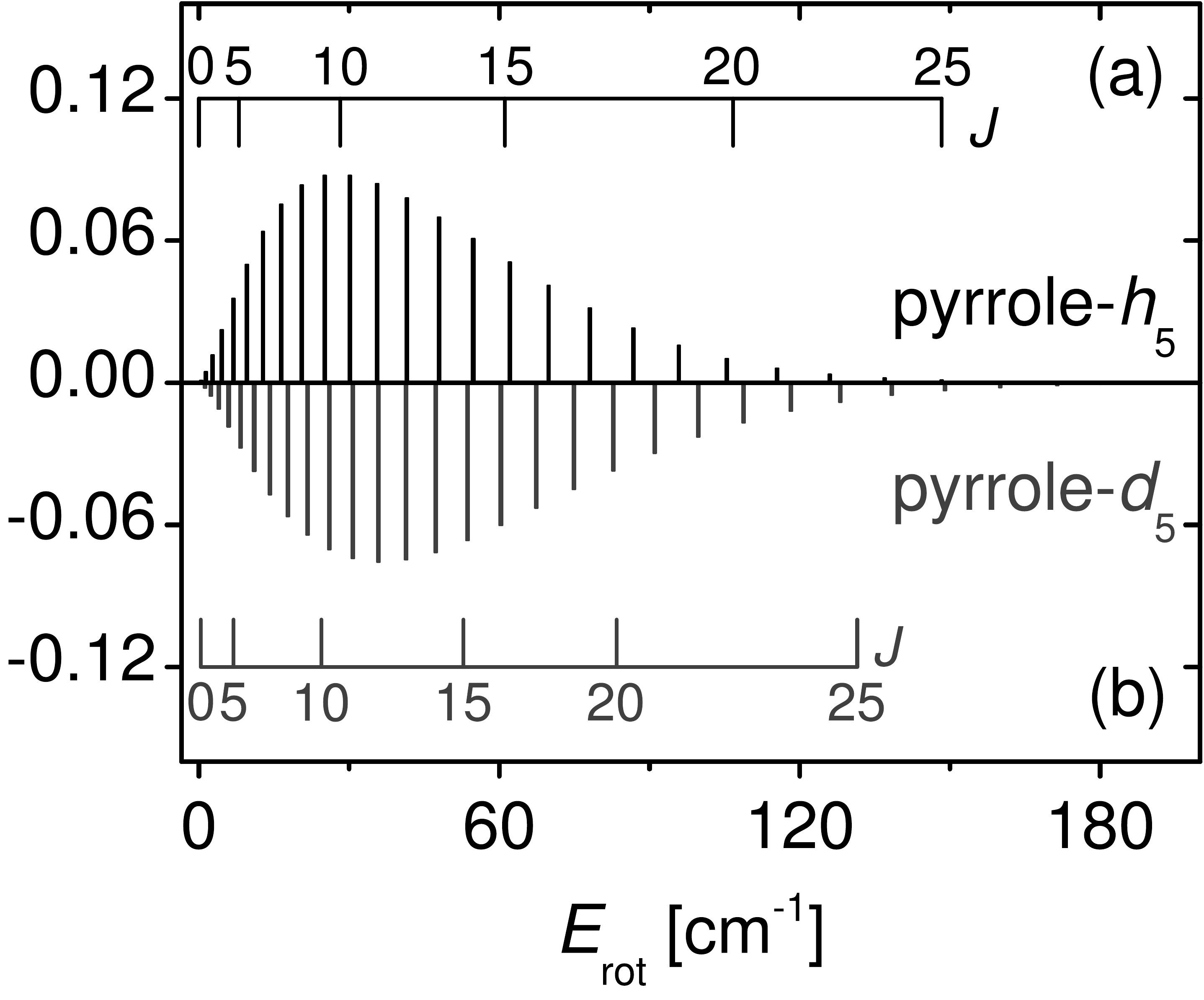}
\caption{Rotational distributions of the pyrrolyl fragment formed upon
 the detachment of the H-atom (a) and D-atom (b) in the state  $1^1\!A_2$
described with the 3D potential of 
the disappearing modes. The TDM function is 
$\mu_y \sim \sin \theta \cos \phi$. 
The distributions are shown versus  the 
angular momentum $j$ and  the rotational
energy $E_{\rm rot}$ of pyrrolyl.
The photon energy is equal to the absorption maximum
of the spectral function $\bar{\sigma}_R$ ($\approx 4.1$\,eV). 
} 
\label{rotdist}
\end{figure}

The individual states in the 
TKER spectra have translational energies $E_{\rm kin} = 
\Eph - E_{\mathbf{n},j,k}$ where $E_{\mathbf{n},j,k}$ is the internal energy
of pyrrolyl which is measured with respect to the ground vibrational 
state in $\widetilde{X}$ and includes the energy of the final rotational state 
$(j,k)$.\cite{NOTE-PYR01B-01} Example of the rotational state distribution,
calculated for the 3D potential of the disappearing modes, is shown in 
Fig.\ \ref{rotdist}(a). The rotational excitation is at best modest: The 
rotational states of pyrrolyl with $j \le 25$ are 
populated. The reason for this is the weak torque along $\theta$ experienced
by the dissociating wave packet. 
The maximum is reached at $j = 10$, corresponding to the rotational
energy of merely 30.6\,cm$^{-1}$, and the full width at half maximum (FWHM) is
near 60\,cm$^{-1}$. This is in agreement with the observation of Cronin et
al. that the product C$_4$H$_4$N is formed rotationally cold.\cite{CNQA04} 
In the vibrational distributions discussed below, the
rotational structure is not resolved: The 
$\delta$-function in Eq. (\ref{tker1}) is
replaced with a Gaussian with a standard deviation of 10\,${\rm cm}^{-1}$, 
and the narrowly spaced rotational states are merged into a single 
peak profile. 

\subsubsection{6D product state distributions: Coordinates 
$R,\theta,\phi,Q_{b_1}(1,2,3)$}
\label{res_tker_1}
In this calculation the $^1\!A_2 \leftarrow \tilde{X}$ excitation is induced 
by the TDM components $\mu_x$ and $\mu_y$ (see paper I for the detailed 
discussion). 
The $x$-polarized transition 
leads to the formation of pyrrolyl in the vibrational states belonging to 
the irrep $a_1$ (associated with rotational states of $b_2$ symmetry).
The $y$-polarized transition excites either the disappearing 
out-of-plane H-atom bending  (irrep $b_1$) or the ring
normal modes of $b_1$ symmetry. If the disappearing bending is excited,
pyrrolyl is formed in the 
$a_1$ vibrational states and $b_1$ rotational states. If the ring modes are
excited, pyrrolyl is formed
in the $b_1$ vibrational states and $a_1$ rotational states.
These symmetry considerations are useful and 
straightforward. In our calculations, they hold because the calculations
are performed for the zero total angular momentum in a single electronic
state and because the 
harmonic Hamiltonian for the $\QQ$ space, based on the block diagonal Hessian, 
prevents the energy exchange between the degrees of freedom 
belonging to different irreps. 

The TKER distributions calculated using the MCTDH and the mapping
approach are illustrated in 
Fig. \ref{F06res} for the excitation energies 
$\Eph = 4.1$\,eV and 
$\Eph= 4.3$\,eV, corresponding to the maximum and the shoulder of the 
spectrum in Fig. \ref{spec}(a).
The TKER spectra, averaged over the field polarizations, 
consist of three peaks. Their rotational width, 
FWHM\,$\approx 60\,{\rm cm}^{-1}$, is the same as for the 
distribution calculated for the 3D case and shown in Fig. \ref{rotdist}(a). 
For the case of MCTDH, the
rovibrational populations are calculated directly. 
For the case of mapping, the
rovibrational populations are evaluated as $|t_{\bm n}(E_{\rm ph})|^2 P_{j,k}$,
with the unity normalized rotational populations $P_{j,k}$ 
calculated for the dissociation in the $\RR$ space at the maximum of the
$\bar{\sigma}_R$ profile.

The assignment of the peaks is easier to clarify first 
using the mapping calculation.
Indeed, from the convolution calculation of the absorption spectrum, we know
the assignment of each absorption band in terms of the ring excitations in
the FC zone (see paper I). Moreover, the mapping of each such ring excitation 
onto free pyrrolyl is also known and 
Fig.\ \ref{adcurves}(c), showing the adiabatic frequencies of the $b_1$
ring modes as functions of $R$, provides an illustration of this mapping.
For $\Eph = 4.1$\,eV
[Fig. \ref{F06res}(a)], the fastest peak corresponds to the 
vibrational ground state ${\bf 0}$ (labeled {\it 1} in the combs in
Fig. \ref{F06res}), and the two slower peaks are the 
fundamental excitations of the modes $Q_{b_1}(2)$
($n_{b1}(2) = 1$; label {\it 2} )
and $Q_{b_1}(3)$ ($n_{b1}(3) = 1$; label {\it 3})
excited by the TDM component $\mu_y$.
The intensity of the mode $Q_{b_1}(1)$ is negligible
because the corresponding Herzberg-Teller
coefficient in the TDM expansion is small. The TKER calculated using MCTDH has
very similar intensities on the same pyrrolyl states.
For the 6D potential $V^{A_2}$, involving only
symmetry breaking ring modes, the dynamics is
vibrationally adiabatic. The adiabatic frequency curves in 
 Fig.\ \ref{adcurves}(c) confirm this conclusion: They are well separated 
and free from avoided crossings.  
Therefore the initial vibrational state distribution of the ring vibrations, 
created during the optical excitation, is carried over to the fragments 
without much energy redistribution.

In their experimental study, Ashfold and co-workers argued that the 
significant intensity observed for the ground vibrational state of pyrrolyl 
was due to vibrationally non-adiabatic effects.\cite{CNQA04} The present 
calculations show that this state is formed because the 
TDM components $\mu_x$ and $\mu_y$ excite the in- and out-of-plane 
bending modes of the detaching H-atom which evolve adiabatically into the 
free rotations of vibrationless pyrrolyl.  
For the higher photon energy, 
$\Eph= 4.3$\,eV [Fig. \ref{F06res}(b)], the TKER
peaks shift to higher kinetic energies but the pattern remains the same.
This is in agreement with the assignment\cite{PG17A}  
of the shoulder band of the 6D
absorption spectrum to the same ring states as the main absorption band
plus one quantum of the NH stretch excitation, 
$n_R = 1$. This excitation
corresponds to a disappearing mode, and leaves no peaks in the TKER. The
color of the assignment combs is meant to 
stress this observation: For the peaks originating
from absorption bands with a given ring excitation and the lowest possible
excitation in the disappearing modes ($n_R =  0$ in this case), the combs
are shown red; for the peaks stemming from the absorption bands with the same
ring excitation plus an additional excitation of the disappearing modes
($n_R = 1$), the combs are blue.

The TKER peaks corresponding to a given pyrrolyl excitation and observed
over a broad range of photon energies are repeatedly found in the 
calculations described below. The peaks merely 
shift to higher kinetic energies with growing $E_{\rm ph}$ but do not 
disappear. The origin of this observation is related to the characteristic
excitation pattern in the absorption spectra discussed 
in paper I: Many absorption bands above the ground state in the local
minimum of the state $1^1\!A_2(\pi\sigma^*)$ carry multiple assignments
which include ring excitations already observed in the low lying bands 
augmented with additional quanta in the disappearing modes.\cite{PG17A}

\subsubsection{11D product state distributions:
  Coordinates $R,\theta,\phi,Q_{a_1}(1,2,3,4,5,6,7,8)$}
\label{res_tker_2}

In the 11D calculation, two TDM components $\mu_x$ and $\mu_y$ mediate the
$^1\!A_2 \leftarrow \tilde{X}$ transition; two TKER spectra 
$I_{{\rm TKER},x,y}(E_{\rm kin}| E_{\rm ph})$ are observed while the 
component $I_{{\rm TKER},z}(E_{\rm kin}| E_{\rm ph})$ vanishes. 
The transition excites exclusively the in- and
out-of-plane H bending modes. Since the initially excited coordinates are
disappearing modes, pyrrolyl is formed in the vibrational states of $a_1$
symmetry (including the ground vibrational state ${\bf 0}$),
associated with rotational states of either $b_2$ or $b_1$ symmetry.

The polarization averaged TKER profiles, 
$$I_{\rm TKER}(E_{\rm kin}|E_{\rm ph}) = \frac{1}{3}
\big[I_{{\rm TKER},x}(E_{\rm kin}| E_{\rm ph}) + 
I_{{\rm TKER},y}(E_{\rm kin}| E_{\rm ph}) + 
I_{{\rm TKER},z}(E_{\rm kin}| E_{\rm ph})\big]\, ,$$ 
corresponding to the specific
absorption bands in the spectrum in Fig. \ref{spec}(b), 
are plotted in Fig. \ref{F07res} for the excitation
energies of $\Eph = 4.40\,{\rm eV}$, $\Eph = 4.65\,{\rm eV}$, and
$\Eph = 4.80\,{\rm eV}$ (diffuse bands A, D and F).
A number of vibrational peaks are visible in each TKER spectrum.
Their rotational line shapes are very similar to the 6D case.

The vibrational states of pyrrolyl, populated in the TKER spectra, clearly
change with the photon energy. As one moves from the absorption band 
A to D to F, the
spectral patterns in TKER become noticeably richer. This is different from the
6D case, and in agreement with our understanding that the consecutive
bands in the absorption spectrum in  Fig. \ref{spec}(b) are dominated by
excitations of different ring modes. Note however that the spectral 
patterns get \lq enriched' rather than replaced. Of course, 
the TKER spectrum calculated for a given absorption band
carries excitations unique to this band. However, the excitations
already observed at lower energies are still visible albeit with lower
intensity. 
The inheritance of the spectral patterns is related to the 
excitations of the disappearing modes in the FC zone. 
The assignments of the major TKER peaks, marked with combs
in Fig. \ref{F07res} and consecutively labeled through all three panels, 
are summarized in Table \ref{T01}.

\begin{figure}
\includegraphics[scale=0.5]{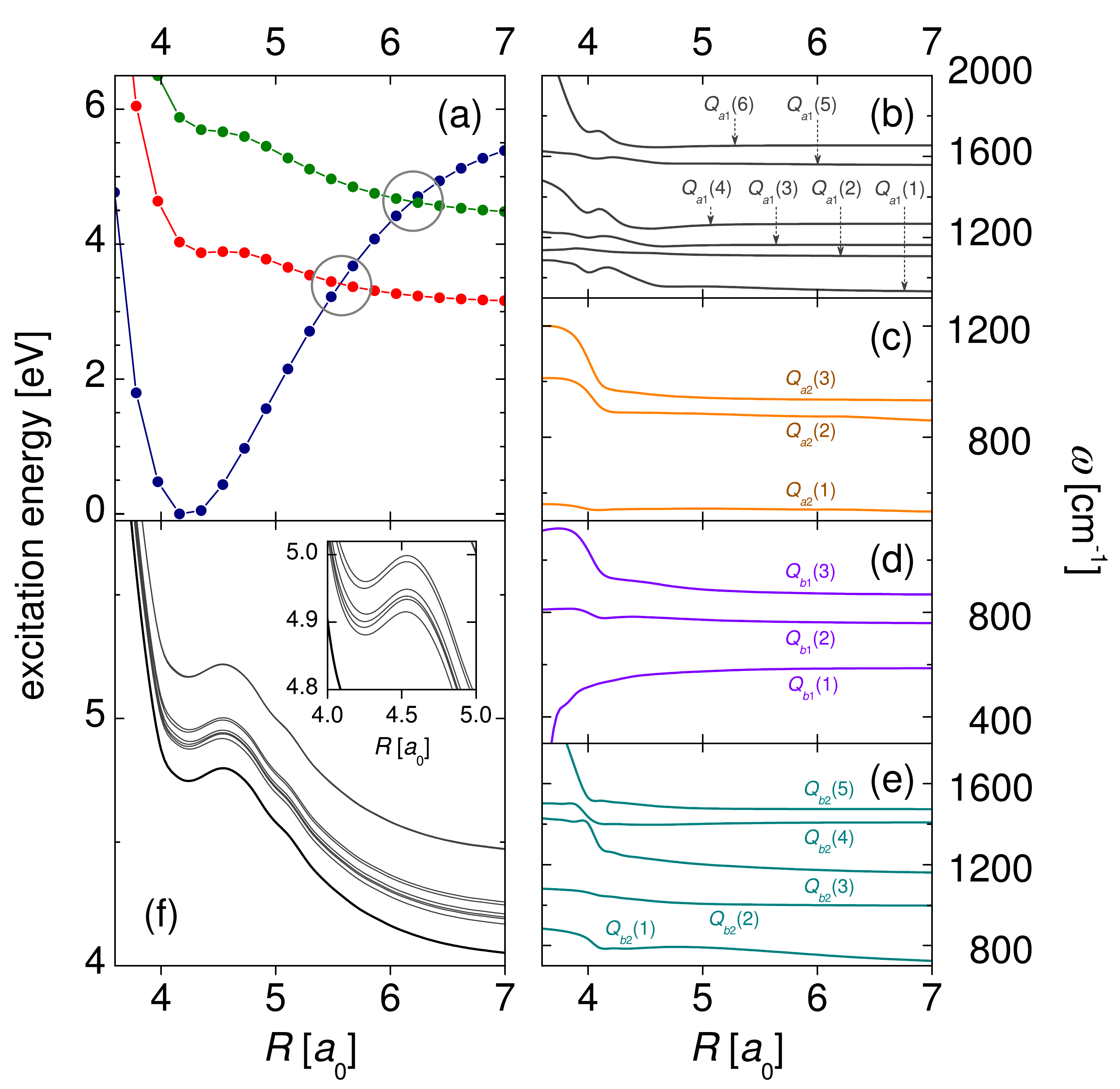}
\caption{
(a) One-dimensional potential energy cuts of the $\widetilde{X}$ (blue),
  $^1\!A_2$ 
  (red), $^1\!B_1$ (green) electronic states as functions of the pyrrolyl-H
  Jacobi distance $R$. Circles mark the conical intersections. 
  (b---e) Frequencies of the normal modes of the ring
  of $a_1$, $a_2$, $b_1$, and $b_2$ symmetry.
  as functions of $R$ 
(f) Adiabatic curves $E_m(R)$ constructed as
described in Sect.\ \ref{map1}. The thick black line is the
zero-point level, the gray lines represent the states with one quantum on the
modes $Q_{a_1}(1,...,8)$. 
}
\label{adcurves}
\end{figure}

The excitation energy of $\Eph = 4.40\,{\rm eV}$ [panel (a)] is close to the 
spectral origin, i.e. the absorption band A. This band is 
due to the excitations of the disappearing bending modes, has zero 
NH stretch excitation ($n_R=0$), and the 
TKER obtained via mapping 
consists of a single rotationally broadened peak of the ground vibrational
state $\bf{0}$ of pyrrolyl (label {\it 1}; cf. Table \ref{T01}). 
This is also the dominant peak in the TKER
distribution calculated using MCTDH. 
However, the photodissociation is not strictly vibrationally
adiabatic, and another state with a one quantum excitation of the low
frequency mode $Q_{a_1}(1)$, $n_{a1}(1) = 1$, is also populated in the quantum
mechanical TKER (label {\it 2}); this peak is not linked to an absorption band
in Table \ref{T01}. There are two clear indications of the vibrational
non-adiabaticity of the mode  
$Q_{a_1}(1)$. While not much displaced in the FC zone, it is the most 
displaced mode between pyrrole and pyrrolyl. As a consequence, 
the $Q_{a_1}(1)$ mode develops an excitation along the dissociation pathway,
away from the FC zone. Second, its frequency plotted against the NH 
stretching  mode $R$ in Fig. \ref{adcurves}(b) shows an avoided crossing
with the adiabatic frequency curve for the mode $Q_{a_1}(2)$ near 
$ R \sim 4.2\,a_0$. 
The states with excitations residing on the higher frequency 
modes have low initial intensities, and their populations are negligible.

The energy $\Eph = 4.65\,{\rm eV}$ corresponds to the absorption band D, and
the TKER spectra are substantially enriched [panel (b)]. 
In the TKER obtained via mapping,
the dominant peaks are {\it 3}, {\it 4}, and {\it 6}, as well as 
{\it 10} and {\it 11}. As indicated in Table \ref{T01}, the first three are
one quantum excitations of the pyrrolyl modes 
$Q_{a_1}(2)$, $Q_{a_1}(3)$, and $Q_{a_1}(5)$. The last two are due to  a
two-quantum excitation and a combination state. The
ring excitation $n_{a1}(5) = 1$, augmented with the obligatory 
fundamental bending 
excitations and $n_R = 0$, is one of the main contributors to the 
absorption peak D. The two other ring states, $n_{a1}(2) = 1$ and 
$n_{a1}(3) = 1$, if combined with the obligatory bending excitations
stem from the absorption band B lying $\sim 0.14$\,eV below the band D.
As discussed in paper I, the ring states are \lq transposed' to the higher
lying absorption 
band via the additional excitation of the disappearing NH bending 
($n_\theta = 3$) in the FC zone. 
Note that 
even in the \lq simple' case of the adiabatic mapping, the assignment of the 
TKER spectra in 11D is complicated, and several additional minor peaks can 
actually be resolved. Strictly speaking, all corresponding ring excitations 
could have been listed as (minor) components of the absorption band D. 

The quantum mechanical TKER are similar in many respects, and most states
predicted in the mapping calculations are populated, although not always
with the same intensities. For the
peaks {\it 3} and {\it 6}, the MCTDH intensities are similar to those
predicted by mapping, while the MCTDH intensities of the peaks 
{\it 10} and {\it 11} are noticeably lower. The discrepancies between MCTDH
and mapping  are
indicative of the vibrationally non-adiabatic photodissociation. A case in 
point is provided by the peaks {\it 4} ($n_{a1}(3) = 1$; absent in the
quantum mechanical TKER) and {\it 5}
($n_{a1}(4) = 1$; absent in the mapping TKER). The ring modes 
$\widetilde{Q}_{a_1}(4)$ and
$\widetilde{Q}_{a_1}(3)$ undergo a strong Duschinsky mixing  
as the distance C$_4$H$_4$N$\cdots$H grows across the FC zone. This is also
seen in the adiabatic frequency curves for these two modes in 
Fig. \ref{adcurves}(b), which exhibit an avoided crossing for 
$R\approx 4.0\,a_0$ and $4.3\,a_0$. In the mapping calculation, the 
asymptotic pyrrolyl mode $Q_{a_1}(3)$ is mapped only on one mode of this pair,
namely 
$\widetilde{Q}_{a_1}(3)$, and only a part of 
the true TKER distribution is recovered. 

The peaks {\it 1} and {\it 2}, inherited from the absorption band A are 
clearly visible, too. The corresponding combs are shown
with blue color in panel (b). Thus, the ground vibrational state
of the ring contributes to the assignment of the high energy band D. This is
again realized via excitation of the disappearing mode, in this case  
the NH stretch (the 
anharmonic frequency of $\sim 2100$\,cm$^{-1}$), possibly accompanied by an
additional low frequency bending excitation. Such moderate
excitations of the disappearing modes are collectively marked as
$n_R =1$ in Table \ref{T01} and in Fig.\ \ref{F07res}. 

The TKER distribution for $\Eph = 4.80\,{\rm eV}$, corresponding to the 
absorption peak F, illustrates the further development of the above 
trends [panel (c)]. Again, several groups of TKER peaks can be 
distinguished. The peaks  {\it 14 --- 18}, shown with red, correspond to 
$n_R = 0$ and a weak bending excitation. 
The assignment of these peaks, given 
in Table \ref{T01}, shows that all of them are combination bands. Note
that two of these ring states are first encountered in 
the absorption band E located slightly below F. The peaks indicated 
with the blue comb are inherited from the absorption band D.
The corresponding pyrrole states in the FC zone carry the same labels in the 
$\QQ$ space, and an additional excitation in the $\RR$ space symbolically
denoted $n_R = 1$. Finally, the gray comb marks the weak 
contribution of states inherited from the low lying band A 
which in the FC zone possess, along with the ring
excitations shown in Table \ref{T01}, a strong excitation $n_R = 2$ of
the disappearing modes. 

The adiabatic mapping is seen to give reliable predictions of the TKER
distributions across the broad photon energy range. In the case of the $a_1$ 
ring modes, strongly coupled to each other and to the dissociation 
coordinate, the accuracy of the mapping calculations 
is the highest for the one quantum excitations. 
For the large amplitude combination states, the vibrationally non-adiabatic
effects become prominent as the 
effective intramode coupling grows. As illustrated
in Fig. \ref{adcurves}(b), the avoided crossings between adiabatic curves and
the associated intramode mixings are restricted to short 
distances $R \le 4.5\,a_0$. 

\subsubsection{15D product state distributions: Coordinates 
$R,\theta,\phi,Q_{a_1}(1,2,5), Q_{a_2}(1,2,3), Q_{b_1}(1,2,3), Q_{b_2}(1,3,5)$}
\label{res_tker_3}

The normal modes of pyrrolyl included in this 15D calculation are sketched in 
Fig. \ref{pyrrolyl_modes}.
The quantum mechanical TKER spectra are calculated for the 
coupled states $^1\!A_2/\tilde{X}$ and shown 
in Fig. \ref{F08res} for the excitation energies 
$\Eph = 4.35\,{\rm eV}$, $\Eph=4.50\,{\rm eV}$ and $\Eph = 4.65\,{\rm eV}$, 
corresponding to the diffuse bands A, C and D in Fig. \ref{spec}(c). The
assignments of the kinetic energy peaks are listed in Table \ref{T02}. 
The internal conversion $A_2 \rightarrow \tilde{X}$
involves less than 10\% of the initial population,\cite{GP17A} and 
the non-adiabatic two-state effects are localized to narrow vicinities 
of the excitation energies corresponding to 
Fano resonances. Our TKERs are calculated at a set of 
fixed photon energies, i.e. in the same way as the experiments are often 
conducted,\cite{AKMNOS10} and Fano resonances mostly pass unnoticed. 
For this reason, the mapping calculations are performed for the single state
$^1A_2(\pi\sigma^*)$. The impact of Fano interference on the product state
distributions is illustrated in Sect.\ \ref{sum} using PHOFEX spectra.

The $^1\!A_2 \leftarrow \widetilde{X}$ transition is induced by the three TDM 
components, which create three different initial states, and are 
associated with product states of different symmetries. The final state 
symmetry is an attribute which can considerably simplify the assignment of
the TKER peaks. The irrep of the 
  fragment vibrational-rotational states is 
given as the direct product $\Gamma_Q \times \Gamma_{\rm ang}$, where 
$\Gamma_Q$ and $\Gamma_{\rm ang}$ are the irreps of the 
vibrational and rotational eigenfunctions, respectively. For the 
$\mu_x$-induced transition, the symmetry allowed vibrational-rotational 
states have $b_2$ symmetry which we find realized in the 
combinations (i$'$) 
$\left(\Gamma_Q = a_1, \ \Gamma_{\rm ang} = b_2\right)$ and  
(ii) $\left(\Gamma_Q = b_2, \ \Gamma_{\rm ang} = a_1\right)$. 
For the $\mu_y$-induced transition, the total symmetry is $b_1$ which in the
calculations is found as (i$''$) 
$\left(\Gamma_Q = a_1, \ \Gamma_{\rm ang} = b_1\right)$ and
(iii) $\left(\Gamma_Q = b_1, \ \Gamma_{\rm ang} = a_1\right)$. Finally, 
for the $\mu_z$-induced transition the total symmetry is $a_2$ which is 
found realized via the combination (iv) 
$\left(\Gamma_Q = a_2, \ \Gamma_{\rm ang} = a_1\right)$.\cite{NOTE-PYR01B-02}
An example of the TKER distribution broken down into the four symmetry blocks
(i)---(iv) is given in Fig.\ \ref{symtker}. 

For $\Eph = 4.35\,{\rm eV}$ [panel (a)], the TKER spectrum is dominated by 
the pyrrolyl fragments in the ground vibrational state
(Table \ref{T02}). The absorption band A
is primarily 
due to the vibrational states having one quantum of the disappearing bending
excitations. This is clearly reflected in both 
the mapping and the quantum mechanical TKER distributions. The vibrational 
states with no bending excitation and one quanta along the modes 
$\widetilde{Q}_{a2}(2,3)$ and $\widetilde{Q}_{b1}(2,3)$ also make
significant contributions to the absorption band A. These are also found in 
the mapping TKER (which is expected) and in the quantum mechanical TKER (which
indicates that the photodissociation follows the vibrationally adiabatic
mechanism). The 
contributions to the TKER spectra arising from the 
pyrrolyl states belonging to a specific  
irrep of the $C_{\rm 2v}$ point group are illustrated 
in Fig.\ \ref{symtker}. The pyrrolyl states of $a_1$ symmetry
originate from the excitation of the disappearing bending vibrations in
the $x$- and $y$-polarized transitions. Several TKER peaks in 
 Fig. \ref{F08res}, 
Fig.\ \ref{symtker}, and in Table \ref{T02} carry multiple assignments. Such
are, for example, peaks $4$ and $5$. The envelope of the peak $4$ hides
the vibrational states of the symmetry $b_1$ and $a_2$. For the peak
$5$, the contributions stem from the vibrational modes $a_1$ and $a_2$. 
Close to the band origin, such \lq blending' of assignments is rare.
Near the absorption maximum, it 
becomes a rule and an obstacle to an unequivocal assignment. 

The mapping approach is reliable for the 15D case. The peak intensities are 
somewhat underestimated only for the pyrrolyl states involving excitations of 
the $a_1$ modes. The intensities of the states with one quantum excitations 
on non-totally symmetric modes are essentially exact.
 Not described in the mapping approach are the  very 
weak structures appearing at ${\rm TKER \approx 6000 \,cm^{-1}}$. They are 
populated non-adiabatically, involve an extra quantum of the mode  
$Q_{a_1}(1)$,  and are discussed for the next photon energy window
where they are clearly seen.

The TKER spectrum for $\Eph = 4.50\,{\rm eV}$ [Fig. \ref{F08res}(b)]
corresponds to the absorption band C. The spectrum is arranged in a pattern
familiar from the 11D analysis. The blue comb marks peaks already
seen at lower energies. These state can be adiabatically connected to 
the FC excitations of the disappearing modes.
The energy spacing between bands A and C is merely 
0.15\,eV, so it can hardly be related to the excitation of the 
NH stretch. 
Instead, an additional excitations of the disappearing bending modes
(with the frequencies of 620\,cm$^{-1}$ and 1020\,cm$^{-1}$) are adequate
candidates. The peaks arising from the excitation of those
states in the band C in the FC zone 
which have the lowest possible excitation in the disappearing modes
are marked red (cf. Table \ref{T02}). 
New in the 15D case is that the states 
originating from additional excitations in the $\RR$ space
(such as peak $5$ in the MCTDH calculations or peak $6$ in the 
mapping calculations) have the largest intensities 
in the TKER spectrum. From the point of view of the mapping approach,
which is a faithful
representation of the excitation pattern in the FC zone, this indicates that,
for example, the state $n_{b2}(3) = 1$ augmented with an excitation of the
disappearing modes, contributes substantially to the assignment of the band
C, along with the other ring states indicated in Table \ref{T02}. 
In the exact quantum dynamics, the peak intensities are 
further affected by the vibrationally non-adiabatic effects. Indeed, the 
quantum mechanical intensities of the states under the red comb are 
consistently lower than predicted in the mapping calculation, while the 
peaks under the blue comb have higher intensities than in the mapping. The
primary reason for this appears to be the nature of the \lq red' states which,
according to Table \ref{T02}, are all combination excitations involving the 
totally symmetric $a_1$ excitations. The 
influence of non-adiabatic interaction during 
photodissociation  tends to depopulate them at the expense of pure \lq blue'
states which have fewer coupling partners at their disposal. This is 
also seen in the adiabatic curves in Fig.\  \ref{adcurves}(b): 
In the FC zone the local frequencies of the modes $Q_{a_1}(1)$ and 
$Q_{a_1}(2)$ oscillate and experience avoided crossings.

The TKER spectrum for $\Eph = 4.65\,{\rm eV}$ [Fig. \ref{F08res}(c)]
corresponds to the absorption band D and is shown here to illustrate the 
complexity of the TKER spectra which is achieved with the ab initio
Hamiltonian at large photon energies. The number of populated states
greatly increases, and a detailed assignment, although technically possible,
becomes meaningless without a clear necessity. Nevertheless, the mapping
approach remains fairly accurate in this case too, and delivers a correct
overall intensity distribution of the TKER peaks.

\subsection{Ab initio quantum mechanics versus experiment}
\label{qm-exp}

In this Section, 
the theoretical TKER spectra for the 15D Hamiltonian
are compared against the experimental TKER distributions measured by
Cronin et al. at long photolysis wavelengths $\lambda \sim 250$\,nm 
using the Rydberg tagging technique.\cite{CNQA04} 
The experimental resolution in this wavelength range is high enough to resolve
populations of the individual vibrational states of pyrrolyl.
As the absorption spectra of the state $1^1\!A_2(\pi\pi^*)$ has not been
measured,
the experimental TKER spectra become the key spectroscopic observables 
against which the  theoretical models can be validated. 

\begin{figure}
\includegraphics[scale=0.33]{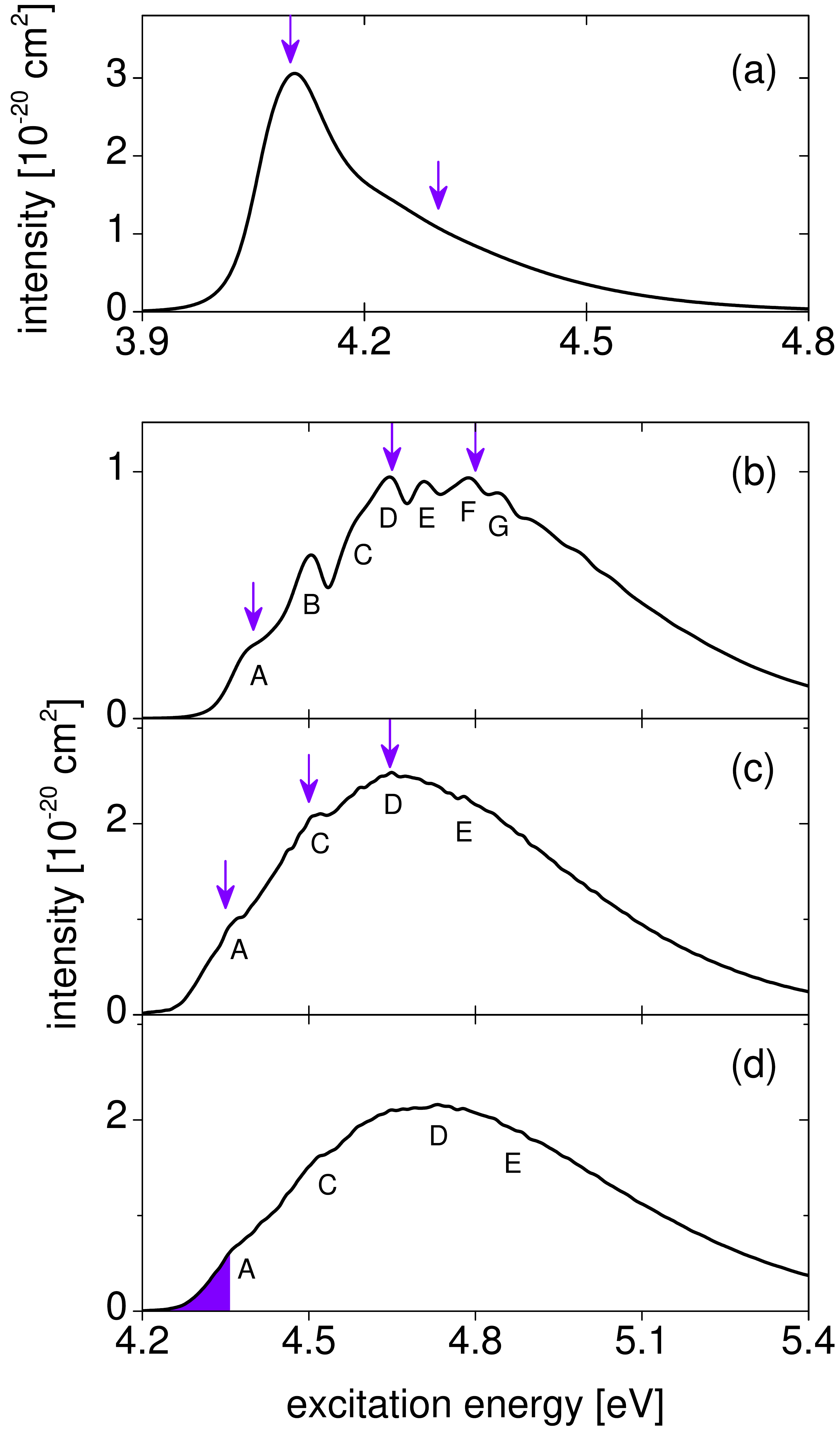} 
\caption{Absorption spectra for the calculations described in Sect.\
\ref{res_tker}: 
(a) 6D; (b) 11D; (c,d) 15D. In (a---c),
the energies, at which the TKER distributions
are analyzed, are marked with arrows. Assignments of the diffuse bands
A---F are discussed in paper I. In (d), the TDM functions 
of Eq.\ (\ref{dipA2_R}) are used, and the area,
in which the comparison with the experiment is made, is shaded.} 
\label{spec}
\end{figure}

\begin{figure}
\includegraphics[scale=0.4]{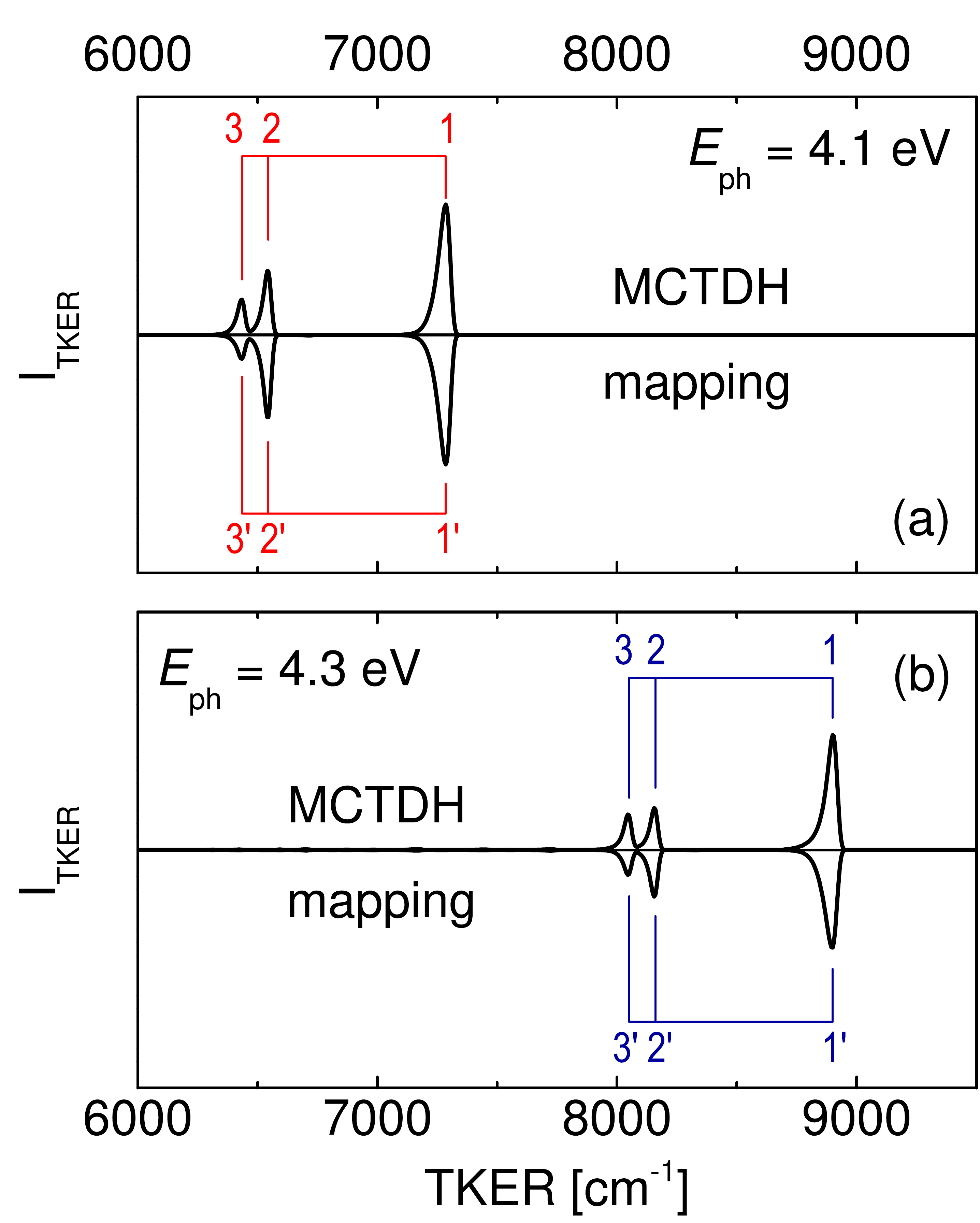}
\caption{TKER distributions calculated for the 
6D spectrum shown in Fig. \ref{spec}(a) for the 
excitation energies
 (a) $\Eph = 4.1\,{\rm eV}$ and (b) 
$\Eph = 4.3\,{\rm eV}$. The quantum mechanical MCTDH distribution 
is shown in the upper, and the distribution calculated via the adiabatic
mapping in the lower half of each panel. The assignments of the peaks
marked with combs: {\it 1.} The ground vibrational state {\bf 0}; 
{\it 2.} $n_{b1}(2) = 1$; {\it 3.} $n_{b1}(3) = 1$. The peaks indicated with 
red combs stem from the vibrational states in the FC zone carrying the
same ring excitation and $n_R= 0$. The peaks indicated with the blue comb
correspond to the vibrational states with $n_R = 1$. 
} 
\label{F06res}
\end{figure}

\begin{figure}
\includegraphics[scale=0.4]{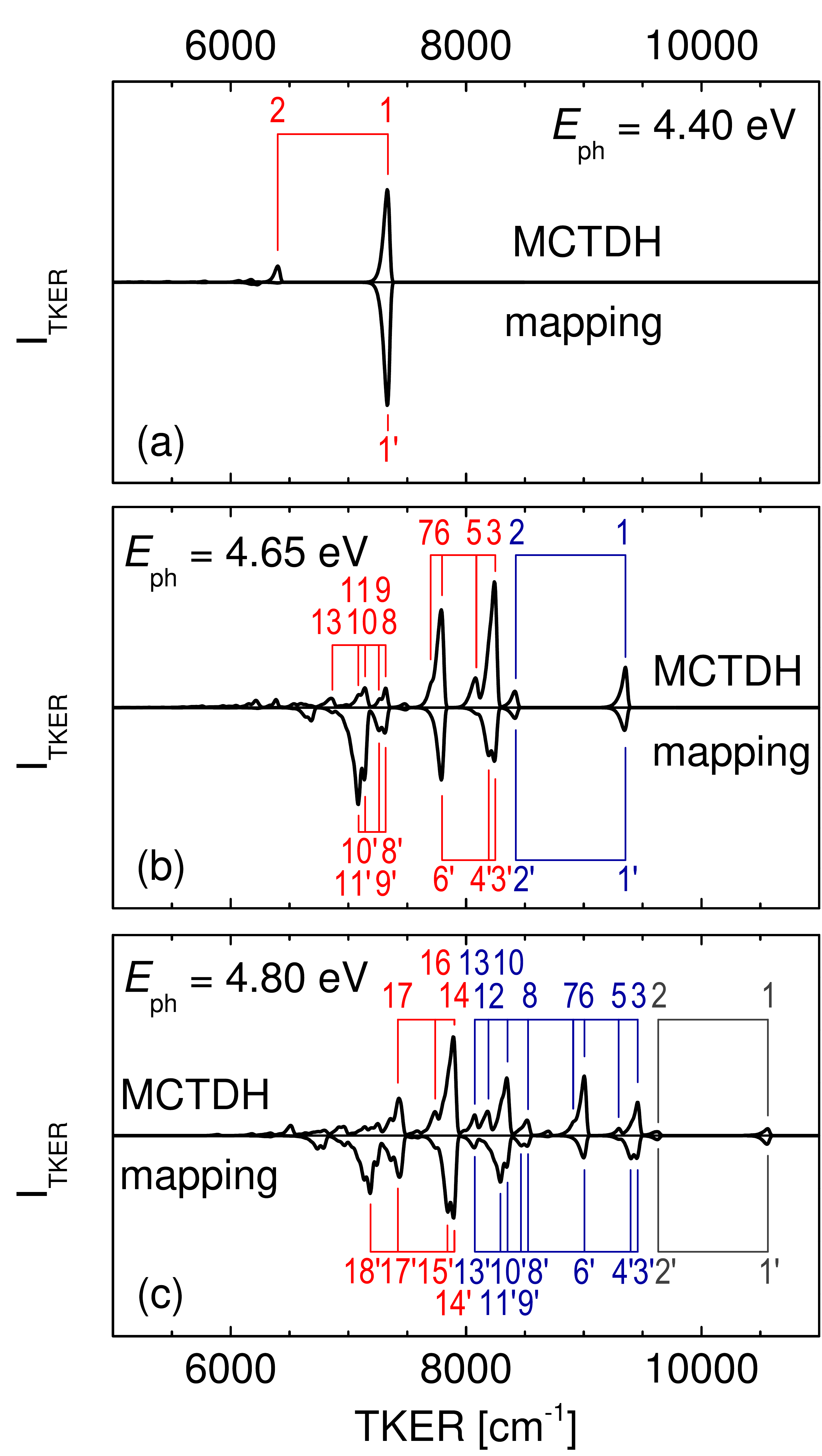} 
\caption{The polarization averaged TKER spectra 
calculated for the 11D spectrum of Fig. \ref{spec}(b) at the excitation 
energies $\Eph = 4.4\,{\rm eV}$ (a), $\Eph = 4.6\,{\rm eV}$(b), 
and $\Eph = 4.8\,{\rm eV}$ (c). The layout follows that of Fig.\
\ref{F06res}. The assignment combs are colored 
according to the extent of excitation of the disappearing modes
in the FC zone: For the fragment peaks originating from the pyrrole states
with $n_R = 0$ and $n_R = 1$, the combs are red and blue; for those
stemming from strongly excited disappearing modes (\lq $n_R = 2$'),
the combs are grey. The assignments are listed in Table\ \ref{T01}.
}
\label{F07res}
\end{figure}

\begin{figure}
\includegraphics[scale=0.4]{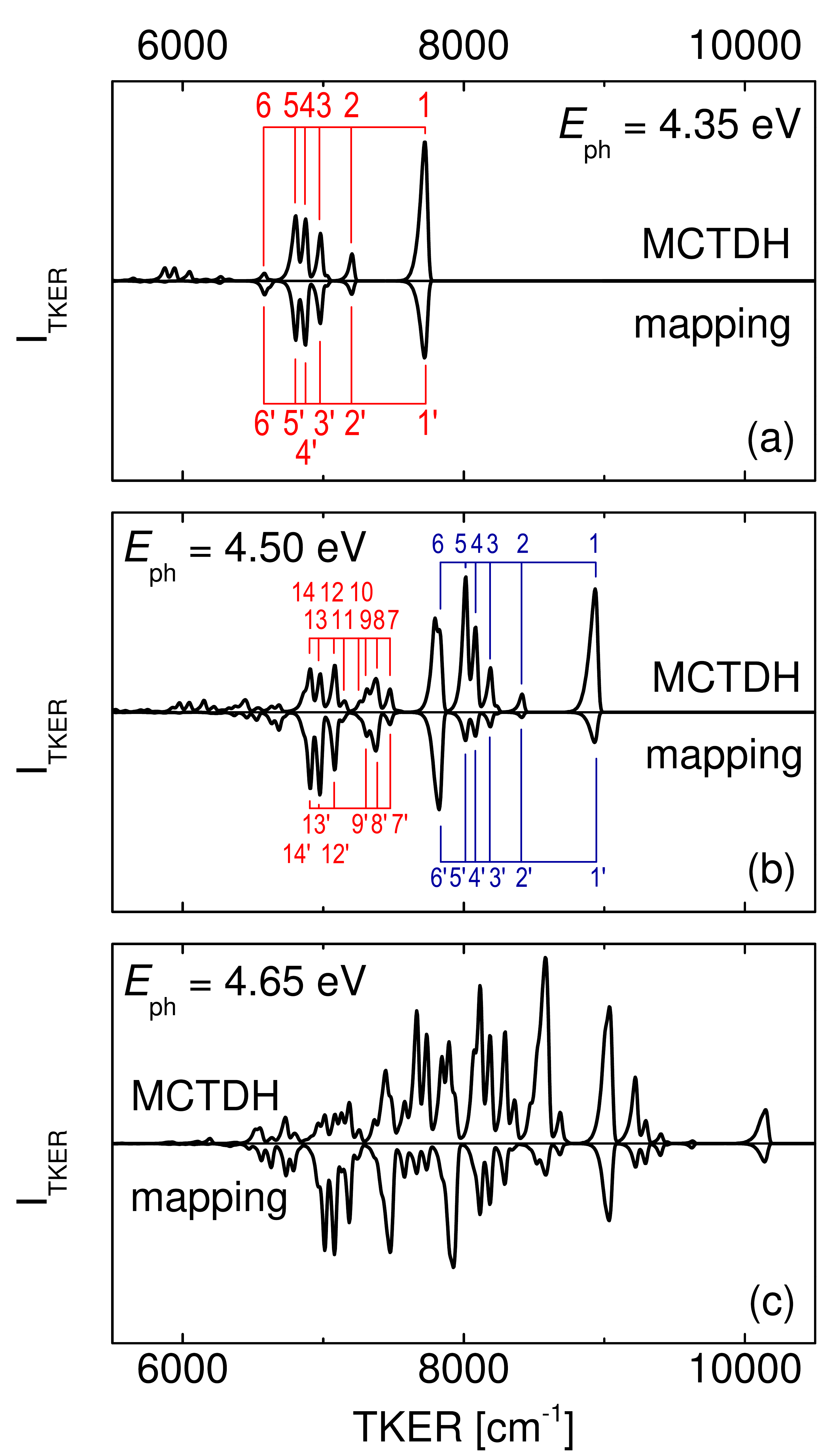} 
\caption{The polarization averaged
TKER distributions calculated for the 15D spectra of Fig. \ref{spec}(c) 
at the excitation energies $\Eph = 4.35\,{\rm eV}$ (a), 
$\Eph = 4.50\,{\rm eV}$ (b), and $\Eph = 4.65\,{\rm eV}$ (c).  The layout
is the same as in Fig.\ \ref{F07res}. 
The assignments are listed in Table\ \ref{T02}.
}
\label{F08res}
\end{figure}

\begin{figure}
\includegraphics[scale=0.5]{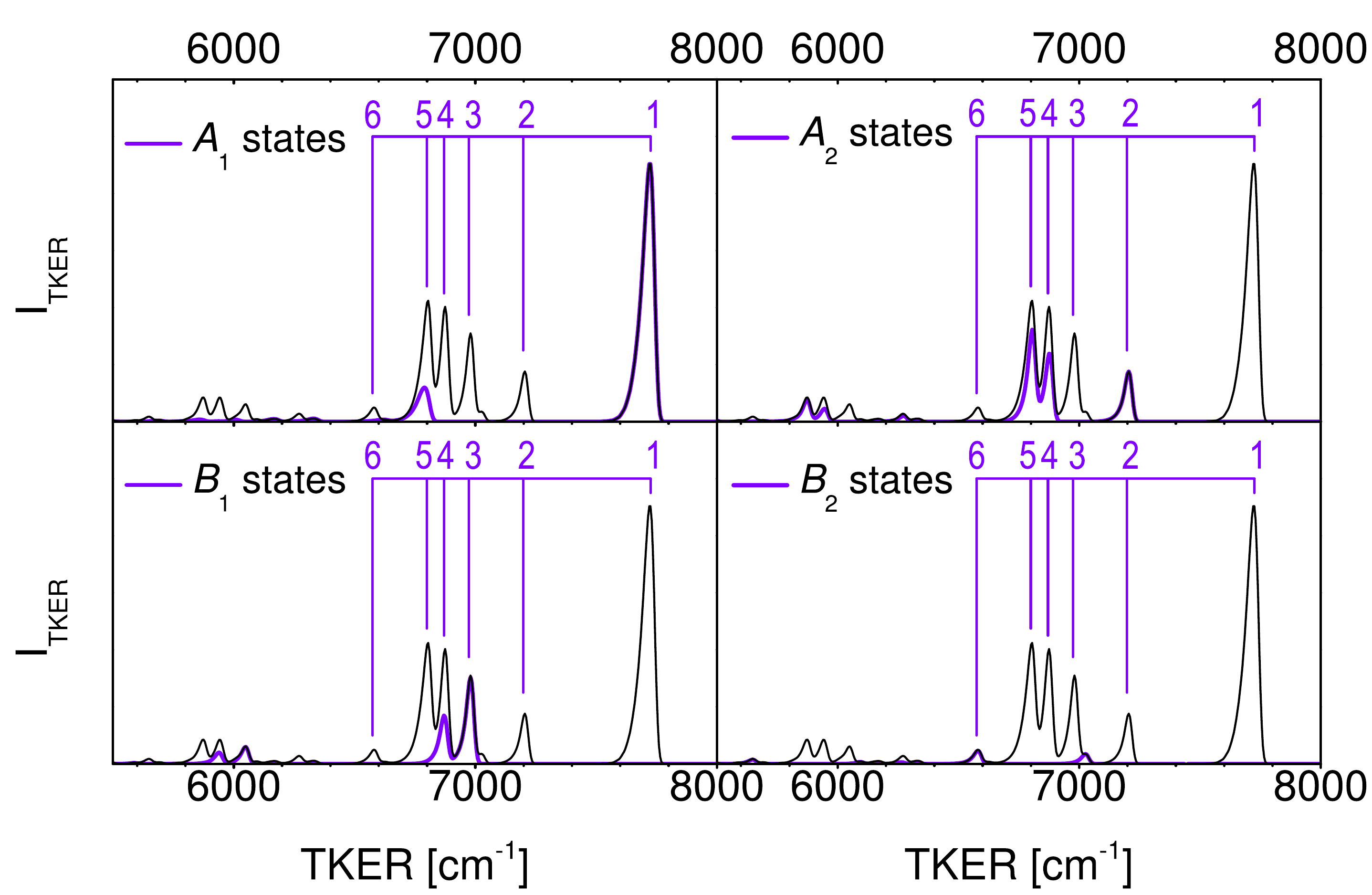} 
\caption{The quantum mechanical 
  TKER distribution calculated for the 15D spectrum of Fig. \ref{spec}(c)
  for $\Eph = 4.35\,{\rm eV}$
(thin gray line), 
decomposed into the contributions corresponding to the 
four symmetry blocks of the $C_{\rm 2v}$ point group (thick
purple lines).
\label{symtker}}
\end{figure}

Pyrrole is described using the same 15 internal 
coordinates of the previous Section 
depicted in Fig.\ \ref{pyrrolyl_modes}. 
In order to provide a realistic representation of the ab initio 
TDMs, the functions of 
Eq.\ (\ref{dipA2_R}) with the $R$-dependent Herzberg-Teller coefficients
are used. The corresponding absorption spectrum is 
shown in Fig.\ \ref{spec}(d). Its overall shape is very similar to the 
absorption spectrum in Fig.\ \ref{spec}(c), calculated for the
same set of coordinates but with the $R$-independent TDM functions. 
The intensity in panel (d) slightly decreases, the spectrum is broadened 
by about $10\%$, while the diffuse vibrational bands are less
pronounced. However, 
the absorption envelope is rather stable with respect to 
changes in the TDM functions because 
the photodissociation is fast and the spectral broadening is substantial. 
More pronounced changes are expected for the 
TKER distributions. Indeed, 
the high degree of vibrational adiabaticity, demonstrated in the previous
sections, implies that the TDM functions in the FC zone can 
significantly influence the product state distributions. 

\begin{figure}
\includegraphics[scale=0.5]{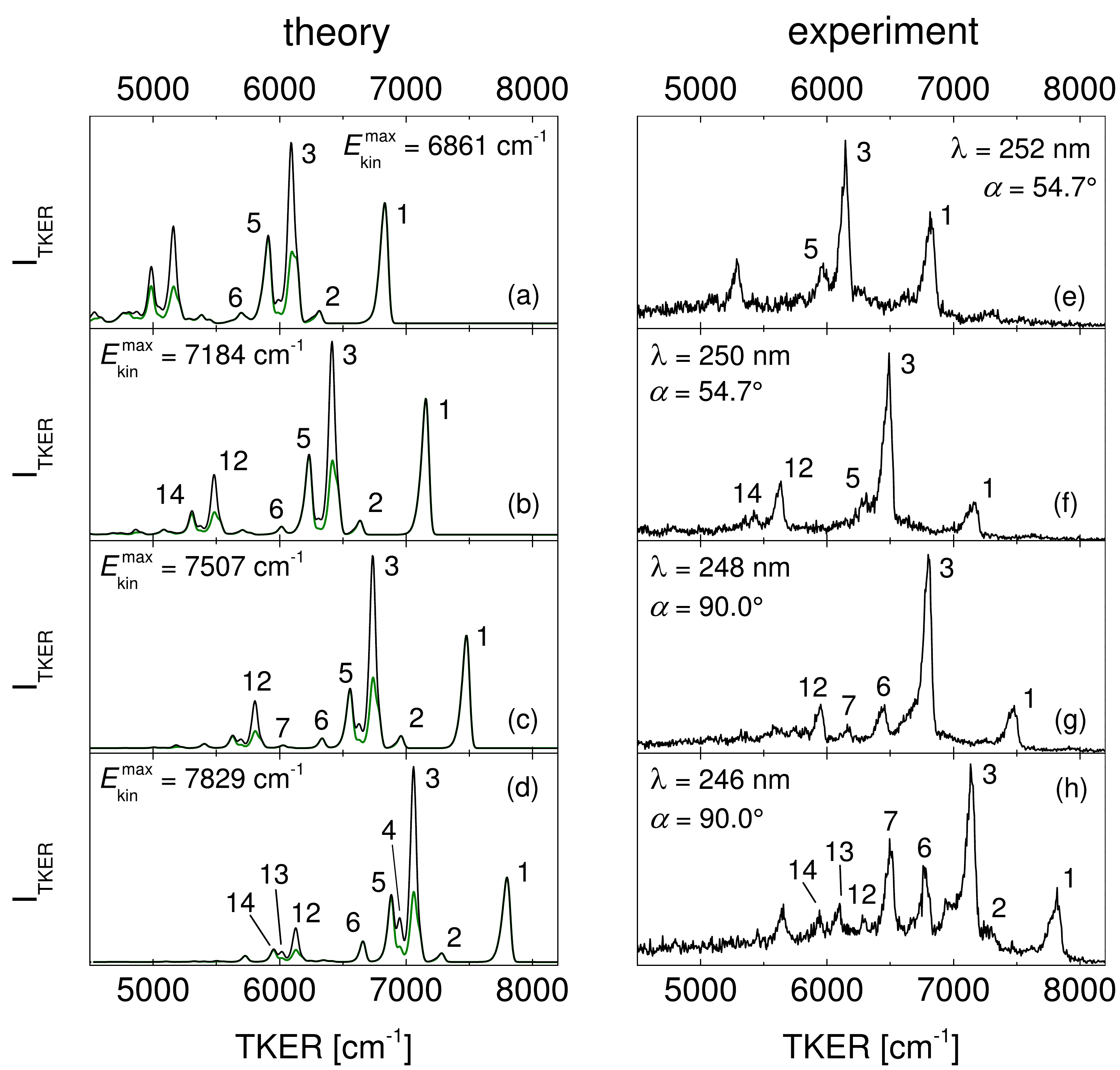}  
\centering
\caption{(a-d) The 15D polarization-averaged TKER spectra calculated with the
$R$-dependent TDM functions (green dashed line). The profiles shown with the
solid black line are obtained by increasing the population of the pyrrolyl
vibrational states of $b_1$ symmetry by a factor three. The kinetic energy
$E_{\rm kin}^{\rm max}$ of the peak corresponding to vibrationless pyrrolyl
is indicated in each panel. (e-h) 
Experimental TKER spectra of Ref.\ \onlinecite{CNQA04}
with similar $E_{\rm kin}^{\rm max}$ as in the
respective panels on the left. The detection angle $\alpha$ and the 
experimental excitation wavelengths are indicated in each panel.
The assignments are listed in Table\ \ref{T02}.
}
\label{F05_a}
\end{figure}

Figure \ref{F05_a} compares the calculated and the measured 
TKER spectra. All results refer to the low energy tail of the absorption 
spectrum highlighted in Fig. \ref{spec}(d). 
The 
polarization averaged distributions $I_{\rm TKER}(E_{\rm kin}|E_{\rm ph})$ 
for the rotationless pyrrole are best compared with the 
TKER profiles measured at the magic angle $\alpha = 54.7^\circ$ 
between the electric field polarization vector and the detection axis
(corresponding to the spectra integrated over $\alpha$). These data are 
published for the two longest wavelengths.\cite{CNQA04} 
For two shorter wavelengths, the
available TKER distributions used in Fig.\ \ref{F05_a}
are for $\alpha = 90^\circ$. 

As explained in 
paper I, the CASPT2 calculations underestimate the 
vertical excitation energy. In order to make a realistic contact with the
experiment of Cronin et al., 
the calculated TKERs are compared with the experimental 
distributions associated with the same maximum available kinetic energy
$E_{\rm kin}^{\rm max}$,
i.e. with the kinetic energy of the vibrationless pyrrolyl,
$E_{\rm kin}^{\rm max} = E_\mathrm{ph} - D_0$. The corresponding experimental
photolysis wavelengths $\lambda$ are indicated in the frames 
(e---h). 

The calculated TKER spectra in panels (a---d) are  depicted with 
green lines. 
All polarizations are included, and --- as in the previous section ---
the TKER spectra peak 
at vibrational states belonging to all four irreps of the point group 
$C_{\rm 2v}$. 
The calculated and the measured distributions are similar in many respects. 
For example, the significantly populated states for all wavelengths 
span a narrow kinetic energy window
of $\approx 2000\,{\rm cm^{-1}}$. Next, 
the rotational widths of individual peaks
is approximately 50\,cm$^{-1}$ implying a cold rotational distribution of
pyrrolyl. Figure \ref{rotdist} demonstrates 
that the rotational excitation of the radical generated upon
dissociation of pyrrole$-d_5$ is stronger than 
for the case of pyrrole$-h_5$.\cite{NOTE-PYR01B-03} 
Finally, three groups of peaks can be clearly recognized in 
the TKER spectra, especially at large wavelengths. With decreasing 
$\lambda$, additional peaks become visible in the measured spectra, but
the original three groups are still visible. 

In principle, the green curves in Fig.\ \ref{F05_a}(a---d)
are already sufficient for a line-by-line comparison with experiment and 
the assignment of the vibrational peaks. However, 
the inspection of the TKER profiles for each polarization separately,  
as is done for example in Fig.\ \ref{symtker}, 
suggests that the populations of the 
pyrrolyl states of $b_1$ symmetry are underestimated in the calculation. 
In order to facilitate the comparison with experiment, we artificially increase
the intensity of the $b_1$ vibrational peaks by a factor three. This 
adjustment gives 
the TKER spectra shown with black lines in panels (a---d). 
The possible origins of the low population of the 
$b_1$ vibrational states in the calculation are discussed in the next section.

The structural similarity between the experimental and the calculated TKERs
permits a dynamics based assignment of the experimental spectra.
The assignment is
easier to summarize for the shortest wavelength shown in Fig.\ 
 \ref{F05_a}(d). In the calculation, this TKER spectrum
is very close to the absorption band A and is similar to the 
TKER spectrum in Fig.\ \ref{F08res}(a). The 
$R$-dependent TDM functions introduce two major modifications into the TKER 
of Fig.\ \ref{F05_a}(d) as compared to 
Fig.\ \ref{F08res}(a): The intensity of the peak corresponding to the 
ground vibrational state of pyrrolyl is decreased, while the intensity of the
peaks around $E_{\rm kin} = 6000$\,cm$^{-1}$ is noticeably enhanced. 
However, the populated vibrational states are the same, 
the corresponding peaks are marked with the same numbers in both figures, 
and  their assignments are summarized Table \ref{T02}. The 
same labeling is also used in Fig.\ \ref{F05_a}(a---c), and 
the evolution of the TKER spectra with increasing wavelength $\lambda$ can be
easily followed. The experimental peaks in Fig.\ \ref{F05_a}(e---f), 
for which the correlation with the
calculated counterpart is straightforward, are also given the same labels. 

The peak {\it 1} 
corresponds to the ground vibrational state of pyrrolyl and
is seen for all excitation wavelengths. In the quantum mechanical TKER,
its intensity is gradually decreasing as the wavelength shortens and 
$E_{\rm kin}^{\rm max}$ grows. This is due to the 
\lq radial factor' $\bar{\sigma}_R(\Eph-E_{\bm 0})$ which decreases 
with growing photon energy. In experiment, the intensity of the peak {\it 1} 
drops more rapidly than in the calculation which might indicate that the
true molecular potential along the NH bond stretch is steeper than in 
the calculation, leading to a narrower spectrum
$\bar{\sigma}_R(\Eph-E_{\bm 0})$. 

The tiny peak {\it 2}, corresponding in the calculation 
to a one quantum excitation of the 
out-of-plane ring mode $Q_{a2}(1)$, 
can be suspected in the experimental spectrum at 250\,nm [panel (f)] 
and is most conspicuous at 246\,nm [panel (h)]. 
The experimental assignment of this peak to the mode $Q_{b_1}(1)$ 
(referred to as $\nu_{21}$ in Ref. \onlinecite{CNQA04}) is not confirmed in
the calculations: Although the mode $Q_{b_1}(1)$ is included, 
its population is suppressed by the low value of the
Herzberg-Teller coefficient in the TDM function.

The strong peak {\it 3}  
is attributed to the fundamental excitation of the mode
$Q_{b_1}(2)$ ($\omega = 757\,{\rm cm^{-1}}$). This coincides with
the experimental assignment (this mode is referred to as $\nu_{20}$ in 
Ref.\ \onlinecite{CNQA04}). In the calculation, 
the one quantum excitation of the mode $Q_{b_2}(1)$ 
($\omega = 710\,{\rm cm^{-1}}$) also makes a minor contribution to 
the intensity of the peak {\it 3}. 

The peak {\it 4}  is revealed in the calculation only in the vicinity of the 
absorption band A [panels (c) and (d)]. It is assigned to the fundamental
excitation of the mode $Q_{b_1}(3)$, and is found in the experimental
TKER as a shoulder of the peak {\it 3}  at lower kinetic energies for 
$\lambda = 248$\,nm and 246\,nm.

The stronger peak {\it 5}
is clearly seen in all calculated and measured TKERs. 
Its assignment, $n_{a2}(3) = 1$, agrees with the experimental one 
(this mode is referred to as $\nu_{9}$ in
Ref.\ \onlinecite{CNQA04}), but the relative 
intensity in the quantum mechanical TKER is overestimated. Note, however,  
that the population of the $Q_{a_2}(3)$ mode in the experiment is 
shown to be strongly dependent on the detection angle $\alpha$:  
It is large for $\alpha = 0^\circ$ (not shown) and small for 
$\alpha = 90^\circ$. It is suggested that the vibrational states of $a_2$ 
symmetry are associated with large anisotropy parameters.\cite{CNQA04}

Finally,  the third group of peaks, labeled {\it 12 --- 14} , is also 
reproduced in the calculation. We assign these peaks to the combination
states built out of excitations assigned to peaks {\it 3 --- 5} augmented  
with an additional quantum of excitation of the totally
symmetric mode  $Q_{a_1}(1)$. No specific assignment was proposed on the 
experimental side, but the authors of Ref.\ \onlinecite{CNQA04} noted that
combinations \lq involving one quantum of any of 
the lower frequency modes of $a_1$ symmetry' are possible candidates. 
Our calculation confirms this suggestion.

\section{Conclusions}
\label{sum}
This paper analyzes the photofragment kinetic energy distributions of
pyrrolyl + H-atom formed in the photodissociation of pyrrole in
the low-lying $^1\!A_2(\pi \sigma^*)$ state. 
The TKER spectra contain 
complementary and, in fact more precise, information on the fragmentation
process than the broad diffuse absorption spectra. The TKER distributions are 
calculated quantum mechanically 
using the 1new quasi-diabatic potential energy matrix with elements 
quadratic in the normal modes of the pyrrolyl ring as described in paper I. 
It has been demonstrated that the quantum mechanical 
TKER spectra can also be efficiently and accurately
reproduced  using the approximate adiabatic mapping approach. Finally, 
calculated TKER spectra are compared with the experimental results.
The main results of our study are summarized as follows:
\begin{enumerate}

\item The peaks in the calculated TKER spectra correspond to the 
few lowest vibrational levels of the pyrrolyl fragment. Sparse TKER 
distributions are seen in all calculations with 6---15 internal coordinates.
States of different symmetries are populated for different electric field 
polarizations. Photodissociation in the ab initio PES of the state  $^1\!A_2$
is characterized by a high degree of vibrational adiabaticity, and  
the TKER profiles are controlled by the initial wave packet shaped by 
the coordinate dependent transition dipole moment. As a result, 
the most populated states are the ground vibrational level, 
the fundamental excitations of the modes mediating the 
$^1\!A_2 \leftarrow \tilde{X}$ transition, and the states with one quantum 
excitations along the strongly displaced $a_1$ modes.

\item The overlap integral-based adiabatic mapping approximation is 
introduced which generalizes the familiar bound-bound FC spectrum calculations
to the case of molecular photodissociation. The mapping approximation 
accurately reproduces the exact TKER distributions and their dependence on 
the excitation energy. This method, which 
requires only a modest numerical
effort largely independent of the molecular size, is a promising tool for the
analysis of the photodissociation dynamics in large classes of 
model biochromophores in which sparse TKER spectra are observed. 

\item The calculated TKER spectra are in good agreement with the distributions
measured in Ref.\ \onlinecite{CNQA04} at wavelengths $\lambda > 246$\,nm. 
The observed populations are reproduced using the 
TDM functions which explicitly depend on the NH stretching mode and
go beyond the Herzberg-Teller expansion. The correlation between theory 
and experiment is sufficiently accurate to allow definitive assignment of the 
measured TKER spectra near the absorption origin. 

\item This work proposes a state specific view on the photodissociation
of pyrrole and shows its power in explaining the dissociation mechanism.  It
also demonstrates that combining two approximate methods, the 
convolution approach to the absorption spectra\cite{PG17A} 
and the adiabatic mapping
of the photofragment distributions, a reliable state specific description
can be achieved without the construction of 
high dimensional potential energy surfaces. A prerequisite for this is a 
substantial degree of vibrational adiabaticity of the decomposition  
which is usually well aligned with the state specificity. 

\end{enumerate}

The constructed ab initio Hamiltonian for the state 
$^1\!A_2(\pi \sigma^*)$
allows one to reproduce the experimental TKER spectra around 250\,nm 
fairly accurately. Nevertheless, the agreement between theory and experiment
is not perfect. Two sources of the remaining discrepancy can be indicated: 

\noindent 
{\it The accuracy of the quantum chemical calculations. } 
The product state populations are strongly dependent on the 
ring mode frequencies and 
the potential energy profile along the NH stretching mode $R$. The frequencies
of the ring modes, discussed in paper I, are accurate to within 5\% or better
for the ground electronic state and the known frequencies of the excite 
states. However, the vertical excitation energies are systematically
underestimated in the CASPT2 calculations. 
Another source of uncertainty is the quality of the CASSCF TDM function, 
which affects the intensity and the shape of the partial cross sections
proportional to the  energy dependent final state populations. The origins of
the direct correlation between the TDM function and the observed 
photofragment distributions are the short time scale of ${\rm <50\,fs}$,
on which the H-atom dissociates and the vibrational adiabaticity of the
reaction. For most vibrational states, the excitation at long wavelengths
corresponds to the weak tails of the partial cross section envelopes. 
For this reason, even small inaccuracies in the PES topography or in the 
TDM functions can lead to large variations in the TKER spectra. In other
words, the photofragment distributions show a typical threshold behavior:  
A small shift in the relative positions of the partial cross sections of 
different fragment states produces a large difference in the relative 
final populations. Pyrrolyl states of all irreps can contribute
because the TDM components $\mu_x$, $\mu_y$, and $\mu_z$  
have comparable magnitudes but form different fragments states. One reason
for the low population of the $b_1$ vibrational states in the
calculation can be an underestimated 
$R$-dependent Herzberg-Teller coefficient for the component $\mu_y$. 
It is sufficient to increase the magnitude of this coefficient by $\sim 50\%$
to bring the intensity of the  $b_1$ peaks to the observed level.

\noindent 
{\it The functional form of the potential energy surfaces.} 
The constructed potential energy surfaces are quadratic in the ring modes,
so that the energy exchange between the vibrations of different symmetries is
possible only indirectly via the common coupling to the NH stretching mode
$R$. This affects the vibrational patterns in the TKER spectra and 
prevents additional fragment states from being directly populated. Moreover, 
the disappearing angular coordinates are coupled to the pool of the ring
modes only via the dissociation coordinate $R$. This can be another reason
for the low populations calculated for the modes of $b_1$ symmetry. Indeed, 
the 
constructed potentials lack coupling between these modes and the 
out-of-plane H-atom bending and the corresponding bilinear coupling terms 
are neglected. They could have induced an energy exchange between the ring and
the disappearing angular coordinates and affect the populations of the
$b_1$ modes. Finally, one has to keep in mind that the molecular Hamiltonian
which is used in the calculations 
is constructed specifically for the dissociation
channel ${\rm C_4H_4N + H}$. Several other decomposition channels were detected
even at low photon energies, and the competing dissociation pathways might
indirectly affect the relative populations observed in the H-atom detachment
channel.\cite{BNL94,WRKRT04,LRHR04,BPPWMBL10}

\begin{figure}
\includegraphics[scale=0.5]{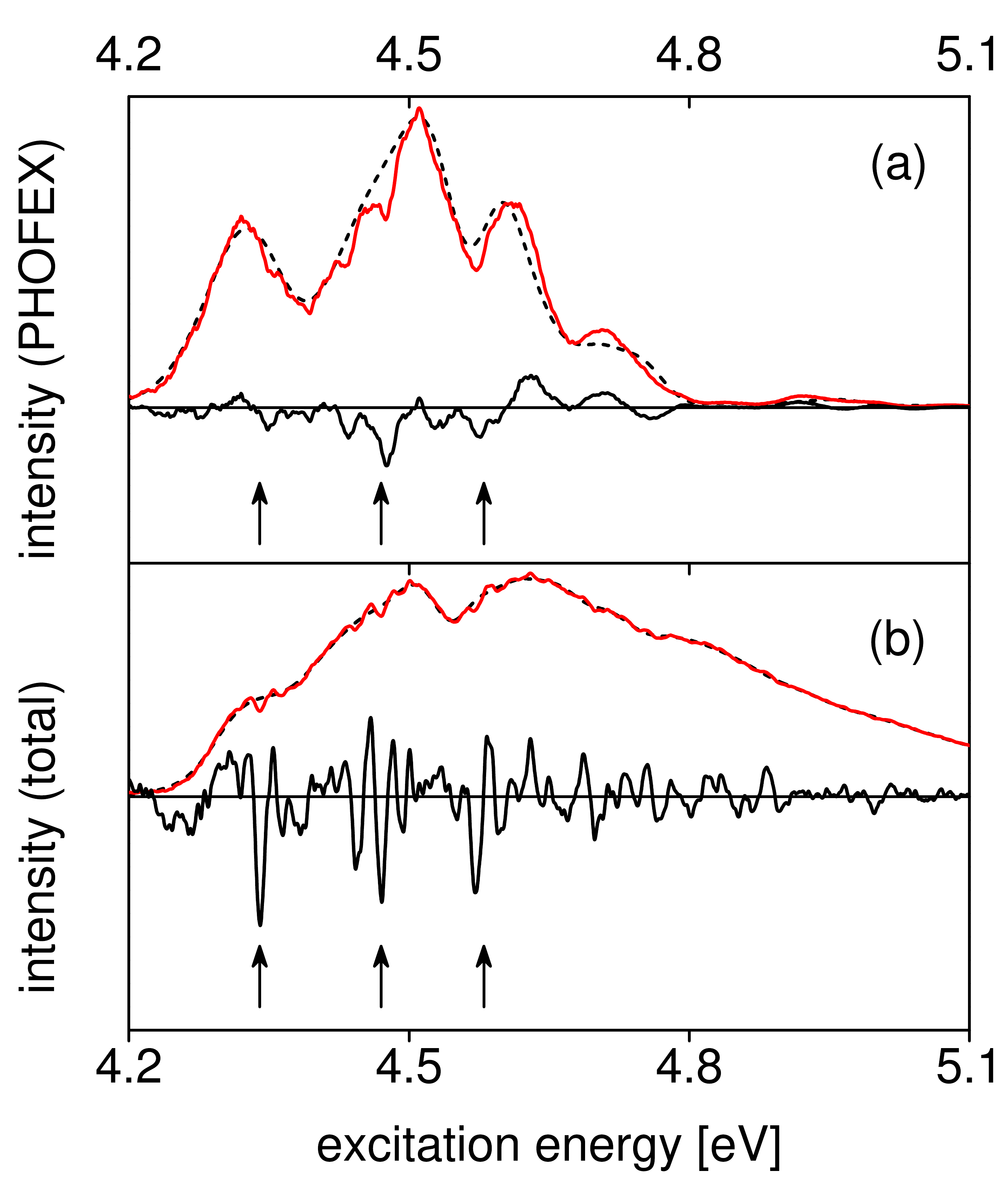} 
\caption{(a) The partial cross section  (the PHOFEX spectrum)
for the
formation of pyrrolyl in the vibrational state with one quantum of the mode
  $Q_{a_1}(1)$ as a function of the photon energy.
The dashed 
black curve is calculated for the single state $1^1\!A_2$; the solid red curve
is for the pair $\widetilde{X}/A_2$. The solid black line is the difference
spectrum. (b) The total absorption spectrum calculated using the 
  TDM $\mu_y \sim \sin \theta \cos \phi$. 
The dashed 
black curve is calculated for the single state $1^1\!A_2$; the solid blue curve
is for the pair $\widetilde{X}/A_2$. The solid black line is the difference
spectrum, multiplied by a factor 10. Arrows in both panels mark the highest
intensity lines in the consecutive multiplets of Fano resonances. 
All calculations are for the 15D case. 
}
\label{phofex}
\end{figure}

We would like to conclude by commenting on the Fano effect which the
conical intersection $\widetilde{X}/A_2$ has on the dissociation
dynamics in the lowest $\pi\sigma^*$ state.\cite{GP17A}. 
In the isolated diabatic state 
$1^1\!A_2(\pi\sigma^*)$, pyrrolyl is formed via a direct dissociation
path closely following, as explained in Sect.\ \ref{res_tker}, the planar
minimum energy path along the coordinate $R$. Upon adding the coupled 
ground electronic
state and the associated CI $\widetilde{X}/A_2$ 
to the picture, the radical ${\rm C}_4{\rm H}_4{\rm N}(^2\!A_2)$ 
 can be formed not only via the direct dissociation of pyrrole, 
\begin{equation}
\label{path1}
\tilde{X}({\bf 0}) \xrightarrow{\bm h\nu}A_2(\pi\sigma^*)
\xrightarrow{\bf diss} {\rm H + pyrrolyl} \, ,
\end{equation}
but also indirectly, with pyrrole making a virtual hop at the intersection,
first to an isoenergetic bound state $|\widetilde{X}({\bf v})\rangle$
and then back to the continuum of $A_2$ 
\begin{equation}
\label{path2}
\tilde{X}({\bf 0})  \xrightarrow{\bm h\nu}  A_2(\pi\sigma^*)
 \xrightarrow{}  \tilde{X}({\bf v})  
\xrightarrow{}\
A_2(\pi\sigma^*)  \xrightarrow{\bf diss}  {\rm H + pyrrolyl} \, .
\end{equation}
The interference between these two diabatic reaction pathways is the origin
of the Fano effect in the photodissociation.\cite{GP17A} In the two state 
absorption spectrum,
this interference results in a series of weak asymmetric Fano peaks 
(\lq ripples') on 
top of the absorption envelope as illustrated in Fig.\ \ref{phofex}(b).  The
resonance structures are enhanced by subtracting the absorption spectrum 
corresponding to the same parent molecular wave function
evolving in the isolated state $1^1\!A_2(\pi\sigma^*)$. This difference 
spectrum, also shown in Fig.\ \ref{phofex}(b), clearly reveals the 
contribution of the optically dark bound vibrational states of the  state
$\widetilde{X}$ to the total absorption spectrum. The resonance lines in the
difference spectrum are
grouped into multiplets. The consecutive
lines with the highest intensity are found
for example near 4.34\,eV, 4.47\,eV, and 4.58\,eV; they are marked with arrows
in Fig.\ \ref{phofex}(b). The spacing between them, 
of the order of 0.12\,eV, matches the frequency of the 
strongly displaced ring mode $Q_{a1}(1)$. The weaker lines making up the
multiplets most probably stem from the anharmonic excitations of the
NH stretch which are strong enough to allow a non-negligible amplitude near
the CI. 

The same interference 
effect also affects the distributions of pyrrolyl over the product vibrational
states. An example is provided in Fig.\ \ref{phofex}(a) for the photon energy
dependent PHOFEX cross section for producing 
pyrrolyl in the vibrational state with a single excitation of the mode
$Q_{a1}(1)$. As in the case of the total absorption spectrum, the Fano 
interference is best revealed in the difference 
spectrum taken between the one- and the two state calculation. 
The resonance 
structures (also marked with arrows)
are correlating well with the multiplets in the absorption spectrum. 
The vibrational structure of the dark bound states in 
$\widetilde{X}$ is carried over to the asymptotic region and imprinted in
the pyrrolyl distributions. At the same time, Fig.\ \ref{phofex}(a) 
demonstrates that the Fano peaks in our calculations are most likely not
fully converged. Many Fano peaks, clearly seen in panel (b), remain unresolved
in panel (a). The reason is the slow population transfer from the state
$\widetilde{X}$ back to $A_2$: Approximately 
$4\%$ of the population still resides in $\widetilde{X}$ 
even after the 1.2\,ps long time evolution. 
Full convergence of Fano resonances requires much longer propagation 
times at which high-dimensional MCTDH calculations can become unreliable.

The Fano resonances in Fig.\ \ref{phofex}
are largely buried in the single state 
background because the ab initio diabatic coupling between the states 
$\widetilde{X}$ and $A_2$ is weak. 
The interstate diabatic coupling is 
much larger at another CI involving the pair 
$\widetilde{X}/B_1$, and for this CI strong Fano interference 
effects are predicted.\cite{GP17A} This topic will be discussed in paper III.

\appendix

\section{Calculation of photofragment distributions with the Heidelberg MCTDH package}
\label{appa}

The projection method of Balint-Kurti\cite{BDM90,BK04} is not implemented
within the Heidelberg MCTDH package. The calculation of the photofragment
distributions is therefore performed
in two steps. In the first step, 
the time-dependent wavefunction 
$\Phi(t) = \exp\left( -i \hat{H}t - \lambda t \right) \Phi(0)$ is
calculated using the MCTDH method and stored at each time step. 
Once the time propagation terminates, 
the $T$-matrix elements are calculated in the second step as follows:
\begin{itemize}
\item For each pyrrolyl eigenstate 
$|\chi_{\bm n}(\QQ)\ket$, the projector $\hat{\cal{P}}_{\bm n}$
of Eq. (\ref{proj1}) is constructed
in the same MCTDH form which is used in the wave packet calculation. 
As the Hamiltonian is set in the normal modes of pyrrolyl, it becomes 
separable as $R \rightarrow \infty$ and the eigenstates 
$|\chi_{\bm n}(\QQ)\ket$ are obtained as single configurations.
\item The time dependent projections of the wave packet on 
each channel state (the cross correlation functions) are calculated as
\begin{equation}
\label{xcorr}
S_{\bm n}(t) =
\left. \left\bra \hat{\cal{P}}_{\bm n}^*\right| \Phi(t) \right\ket\, . 
\end{equation}
This is the time dependent counterpart of the matrix element 
$ \langle\hat{\cal{P}}_{\bm n}^*| \Psi^\lambda(E_{\rm ph})\rangle $ which enters
Eq.\ (\ref{cross4}). 
\item The photon energy-dependent $T$-matrix elements are obtained as 
half-Fourier transforms of the cross-correlation functions,
\begin{equation}
T_{\bm n}(\Eph) \sim \sqrt{\frac{k_{\bm n}}{m_R}} \int_0^\infty S_{\bm n}(t) e^{i \Eph t} dt \ .
\end{equation}
\end{itemize}

\vspace{-0.7cm}
\begin{acknowledgments}
S.Yu.G. acknowledges the financial support by the  Deutsche
  Forschungsgemeinschaft. 
\end{acknowledgments}

\clearpage
\newpage


\newpage

\clearpage
\newpage

\begin{table}

\caption{
Expansion coefficients of the TDM function of Eqs. 
(\ref{dipA2_R}). Values are in atomic units [$e a_0$].
}
\label{T00}
\begin{ruledtabular}
\begin{tabular}{c|ccc}
 & $\mu^{(0)}(R_\mathrm{FC})$ & $\mu^{(1)}(R_\mathrm{FC})$ & $\mu^{(2)}(R_\mathrm{FC})$ \\
\hline 
$\mu^{A_2}_{x,\theta,1} $ & -2.04298  &  -12.2036  &  -9.33618  \\
$\mu^{A_2}_{x,\theta,2} $ & 1.19530  & 6.10336  & 4.35229  \\
$\mu^{A_2}_{x,1} $ & -0.0216261  & -0.0216261 &  -0.0739594 \\
$\mu^{A_2}_{x,3} $ & 0.030926  & 0.0297996  & 0.0018997  \\
$\mu^{A_2}_{x,5} $ & 0.0166159  & -0.0345444  &  -0.0357879 \\
$\mu^{A_2}_{y,\theta,1} $ & 3.05435 & 9.32314 & 19.0464 \\
$\mu^{A_2}_{y,\theta,2} $ & -1.70042  &  -4.37171 &  -9.27847  \\
$\mu^{A_2}_{y,1} $ & 0.00133072  & 0.0186773  &  -0.00116267 \\
$\mu^{A_2}_{y,2} $ & -0.0261693 & -0.0105274  & -0.00296679  \\
$\mu^{A_2}_{y,3} $ & -0.0219039 & 0.0505846 & 0.0469725  \\
$\mu^{A_2}_{z,1} $ & -0.022233 & 0.0534878 & 0.0889894  \\
$\mu^{A_2}_{z,2} $ & 0.0246797  & -0.111224  & -0.154991  \\
$\mu^{A_2}_{z,3} $ & 0.0340891  & 0.0166484  &  0.0265491
\end{tabular}
\end{ruledtabular}
\end{table}

\begin{table}
\caption{
Assignments of the TKER spectra in Fig.\ \ref{F07res} 
in terms of the vibrational states of the pyrrolyl ring
in the 11D calculations. The absorption bands dominated by the same 
ring excitations and the obligatory excitations of the
disappearing modes (\lq $n_R = 0$')
are indicated in the third column. ${\bf 0}$ stands for
the ground vibrational state of the ring. 
}

\label{T01}
\begin{ruledtabular}
\begin{tabular}{ccc}
  &  &   \vspace{-0.5cm} \\
TKER peak & Assignment  & Absorption band \\
\hline
1 & {\bf 0} & A \\
2 & $n_{a1}$(1) = 1 & \\
  &  &   \vspace{-0.5cm} \\
3 & $n_{a1}$(2) = 1 & B \\
4 & $n_{a1}$(3) = 1 & B \\
5 & $n_{a1}$(4) = 1 &  \\
6 & $n_{a1}$(5) = 1 &  C,D \\
7 & $n_{a1}$(6) = 1 &   \\
8 & ($n_{a1}$(1) = 1,$n_{a1}$(2) = 1) &   \\
9 & ($n_{a1}$(1) = 1,$n_{a1}$(3) = 1) &   \\
10 & $n_{a1}$(2) = 2 & D  \\
11 & ($n_{a1}$(2) = 1,$n_{a1}$(3) = 1) &  D \\
12 & ($n_{a1}$(2) = 1,$n_{a1}$(4) = 1) &   \\
13 & ($n_{a1}$(1) = 1,$n_{a1}$(5) = 1) &   \\
14 & ($n_{a1}$(2) = 1,$n_{a1}$(5) = 1) &  E \\
  &  &   \vspace{-0.5cm} \\
15 & ($n_{a1}$(3) = 1,$n_{a1}$(5) = 1) &  E \\
16 & ($n_{a1}$(4) = 1,$n_{a1}$(5) = 1) &   \\
17 & ($n_{a1}$(1) = 1,$n_{a1}$(2) = 2) &  F  \\
18 & ($n_{a1}$(2) = 2,$n_{a1}$(3) = 1) &  F \\
\end{tabular}
\end{ruledtabular}
\end{table}


\begin{table}
\caption{
Assignments of the TKER spectra in Fig.\ \ref{F08res} 
in terms of the vibrational states of the pyrrolyl ring
in the 15D calculations. Contributions of the excitations via the TDM
components $\mu_x$, $\mu_y$, nd $\mu_z$ are shown separately except for 
the states of $a_1$ symmetry which originate from both $\mu_x$ and 
$\mu_y$ components. 
The absorption bands for the $y$-polarized transition, 
dominated by the same 
ring excitations and \lq $n_R = 0$', are indicated in the fifth column. 
} 
\label{T02}
\begin{ruledtabular}
\begin{tabular}{ccccc}
  &  &  & &  \vspace{-0.5cm} \\
TKER peak & \multicolumn{3}{c}{Assignment}  & \makecell{Absorption\\ band} \\
          & $\mu_x$ & $\mu_y$ & $\mu_z$ &  \\
\hline
1 & \multicolumn{2}{c}{{\bf 0}} & & A \\
2 &  &  & $n_{a2}(2) = 1$ & \\
3 &  & $n_{b1}(2) = 1$ & & A \\
4 &  & $n_{b1}(3) = 1$ & $n_{a2}(2) = 1$& A \\
5 & \multicolumn{2}{c}{$n_{a1}(1) = 1$} & $n_{a2}(3) = 1$& A,B \\
6 & $n_{b2}(3) = 1$ & & & \\
  &  &  & &  \vspace{-0.5cm} \\
7 & $n_{b2}(5) = 1$ & & & \\
8 & \multicolumn{2}{c}{$n_{a1}(5) = 1$} &  & C  \\
9 &  &  & ($n_{a1}(2) = 1,n_{a2}(1)=1$)& C \\
10 &  & ($n_{a1}(1)=1,n_{b1}(2) = 1$) & &  \\
11 & & ($n_{a1}(1)=1,n_{b1}(3) = 1$) & ($n_{a1}(1)=1$,$n_{a2}(2) = 1$)&  \\
12 &  & ($n_{a1}(2)=1,n_{b1}(2) = 1$) & & C \\
13 &  & ($n_{a1}(2)=1,n_{b1}(3) = 1$) &($n_{a1}(2)=1,n_{a2}(2) = 1$) & C \\
14 &  &  & ($n_{a1}(2) = 1,n_{a2}(3)=1$)& C \\
\end{tabular}
\end{ruledtabular}
\end{table}

\clearpage
\newpage


\end{document}